\documentclass[12pt, draftclsnofoot, onecolumn]{IEEEtran}
\pdfoutput=1
\ifCLASSINFOpdf
\else
\fi

\usepackage{amsthm,amssymb,graphicx,multirow,amsmath,color,amsfonts}
\usepackage[update,prepend]{epstopdf}
\usepackage[noadjust]{cite}
\usepackage{tabulary}
\usepackage{bbm} 
\usepackage{multirow}
\usepackage{comment}

\def\nb0{{\mathbf{0}}}
\def\nb1{{\mathbf{1}}}

\newtheorem{lemma}{Lemma}

\newtheorem{theorem}{Theorem}

\newtheorem{remark}{Remark}

\def\E{\mathbb{E}}

\allowdisplaybreaks 
\usepackage{setspace}	
\allowdisplaybreaks 
\begin{document}
\title{
Joint Uplink and Downlink Coverage Analysis of Cellular-based RF-powered IoT Network
}
\author{
Mustafa A. Kishk and Harpreet S. Dhillon
\thanks{The authors are with Wireless@VT, Department of ECE, Virginia Tech, Blacksburg, VA (email: \{mkishk, hdhillon\}@vt.edu). The support of the U.S. NSF (Grants CCF-1464293, CNS-1617896, and IIS-1633363) is gratefully acknowledged. This paper was presented in part at the IEEE Globecom, Washington DC, 2016~\cite{conf_version}. \hfill Manuscript last updated: \today.} 
}

\maketitle
\begin{abstract}
Ambient radio frequency (RF) energy harvesting has emerged as a promising solution for powering small devices and sensors in massive Internet of Things (IoT) ecosystem due to its ubiquity and cost efficiency. In this paper, we study joint uplink and downlink coverage of cellular-based ambient RF energy harvesting IoT where the cellular network is assumed to be the only source of RF energy. We consider a time division-based approach for power and information transmission where each time-slot is partitioned into three sub-slots: (i) {\em charging sub-slot} during which the cellular base stations (BSs) act as RF chargers for the IoT devices, which then use the energy harvested in this sub-slot for information transmission and/or reception during the remaining two sub-slots, (ii) {\em downlink sub-slot} during which the IoT device receives information from the associated BS, and (iii) {\em uplink sub-slot} during which the IoT device transmits information to the associated BS. For this setup, we characterize the {\em joint coverage probability}, which is the joint probability of the events that the typical device harvests sufficient energy in the given time slot and is under both uplink and downlink signal-to-interference-plus-noise ratio (SINR) coverage with respect to its associated BS. This metric significantly generalizes the prior art on energy harvesting communications, which usually focused on downlink or uplink coverage separately. The key technical challenge is in handling the correlation between the amount of energy harvested in the charging sub-slot and the information signal quality (SINR) in the downlink and uplink sub-slots. Dominant BS-based approach is developed to derive tight approximation for this joint coverage probability. Several system design insights including comparison with regularly powered IoT network and throughput-optimal slot partitioning are also provided. 

\end{abstract}
\IEEEpeerreviewmaketitle
\section{Introduction}
Internet of Things (IoT) is a massive ecosystem of interconnected {\em things} (referred to as IoT devices) with sensing, processing, and communication capabilities~\cite{whitmore2015internet}. Due to its ubiquity, cellular network has emerged as an attractive option to provide reliable communication infrastructure for supporting and managing these networks~\cite{EriM2015b,DhiHuaJ2013,DhiHuaJ2014,DhiHuaJ2017}. This new communication paradigm will enable a new era of applications including medical applications, transportation, surveillance, and smart homes to name a few. 
Unlike human-operated cellular devices, such as smart phones and tablets, that can be charged at will, these IoT devices may be deployed at hard-to-reach places, such as underground or in the tunnels, which makes it difficult to charge or replace their batteries. This has led to an increasing interest in energy-efficient communication of IoT devices, both from the system design~\cite{DhiHuaJ2013,DhiHuaJ2014,DhiHuaJ2017}, and hardware perspectives~\cite{jelicic2012analytic}. While these efforts will increase the lifetime of these devices, they do not necessarily make them {\em self-sustained} in terms of their energy requirements. One possible way to develop an almost {\em self-perpetuating IoT network} is to complement or even circumvent the use of conventional batteries in the IoT devices by energy harvesting. While one can use any energy harvesting method depending upon the deployment scenario, such as solar energy, thermo-electronic, and mechanical energy~\cite{SomGiaM2015}, we focus on the ambient RF energy harvesting~\cite{Huang2015,6951347}, where the IoT device harvests energy through wireless RF signals. This is because of the ubiquity of RF signals even at hard-to-reach places where the other popular sources, such as solar or wind, may not be available. Besides, RF energy harvesting modules are usually cheaper to implement, which is another consideration in the deployment of IoT devices~\cite{kamalinejad2015wireless}. Now if RF energy harvested from the communication network (cellular network in this case) is the only source of energy, there will obviously be some new design considerations due to the limitations in the energy availability and the correlation in the communication and energy harvesting performance \cite{UluYenJ2015}. In this paper, we concretely expose these design considerations using tools from stochastic geometry. In particular, we define and analyze a new {\em joint coverage probability} metric, which significantly generalizes prior art in this area. Before going into the details of our contributions, we discuss prior art next.%

\subsection{Prior Art}
Owing to their remarkable tractability and realism, tools from stochastic geometry have received significant attention over the past few years for the system-level analysis of cellular networks. Interested readers are advised to refer to \cite{AndGupJ2016,ElSHosJ2013,MukB2014,ElSSulJ2017} and the references therein for a more pedagogical treatment of this topic. More relevant subset of these works for this paper is the one that focuses on characterizing the performance of energy harvesting communication networks; see~\cite{flint2015performance,zhong2015wireless,huang2014enabling,sakr2015analysis,7009965,DhiLiJ2014} for a small subset. In this Subsection, we will discuss these works in the broader context of uplink, downlink, and joint uplink/downlink coverage analyses.



{\em Uplink analysis}. Most of the stochastic geometry-based works in this area are focused on the setups in which the device of interest first harvests ambient RF energy and then transmits information to its designated node (which will be its {\em serving BS} in the uplink cellular network) using this energy. Since the device that harvests energy is also the one that transmits information, we discuss all these works under the category of {\em uplink analysis} to put things in the correct context. The general theme of these works is to study the joint energy and uplink SINR coverage, which is defined as the joint probability of harvesting sufficient ambient RF energy to enable uplink transmission, and having uplink SINR above a predefined threshold. The energy and uplink SINR coverage events are independent by construction if one assumes that the ambient RF sources are placed independently of the communication network~\cite{flint2015performance,zhong2015wireless,huang2014enabling}. A few representative works in this direction are discussed next. Authors in~\cite{flint2015performance} studied a system of energy harvesting wireless sensor network where a sensor node harvests ambient RF energy from the broadcast TV, radio, and cellular signals. The sensor node uses this energy to transmit information to a data sink located at a fixed distance. Authors in~\cite{zhong2015wireless} studied a point-to-point (source-destination) communication link consisting of an energy harvesting source that is powered by a power beacon (PB). In particular, the source harvests power from the RF signals of PB using which it transmits information to its destination. The assumption of the existence of dedicated PBs was then generalized in~\cite{huang2014enabling} which studied the uplink performance of a cellular network in which mobile users are powered by a network of PBs. The other general setup, in which the prior art is significantly sparser, is the one where the same network of BSs is used for charging and communication~\cite{sakr2015analysis,7009965}. This naturally correlates the energy and uplink SINR coverage events. However, to maintain tractability, all prior works study energy and uplink coverage events separately with \cite{sakr2015analysis} justifying it by assuming full channel inversion power control. While such simplifications may work in specific system setups, it is desirable to handle correlation in the two coverage events properly, which will be done as a special case of our analysis.

{\em Downlink analysis}. Another general theme in the literature is to explore setups in which the device of interest first harvests ambient RF energy and then uses it to receive information. We will discuss all these works under the general category of {\em downlink analysis}. In small devices with severely limited power budgets, which is the case for IoT devices, energy consumption during information reception can be almost as important as the energy consumption during uplink transmission. For instance, many recent works have shown that receiver energy consumption scales noticeably with the data rates due to increase in the length of decoder interconnects~\cite{Ganesan2011,blake2015energy,4313073}. Motivated by this general fact, some aspects of system design have already been explored with the consideration of receiver energy consumption, e.g., see~\cite{krikidis2014simultaneous,zhou2013wireless,zhou2015outage} for a subset. For instance, authors in~\cite{krikidis2014simultaneous} used tools from stochastic geometry to study the SINR outage probability and average energy harvested under power splitting at the receiver in a system of randomly placed transmitter-receiver pairs where each transmitter has a unique receiver at a fixed distance. The main objective is to minimize the SINR outage probability subject to a constraint on the minimum average harvested energy. Authors in~\cite{zhou2013wireless} explored power splitting receiver architecture in a point-to-point system to study the tradeoff between the average harvested energy and the average data rate. For this setup, the achievable rate-energy regions are also derived for different types of receiver architectures. Finally,~\cite{zhou2015outage} explored power control policies for outage minimization in a point-to-point link assuming energy harvesting at both the transmitter and the receiver. The outage is said to occur if the signal-to-noise ratio (SNR) is low or the energy harvested at the transmitter or receiver is not high enough. Contrary to all these works, which are more applicable to ad hoc or decentralized networks, the joint analysis of harvested energy and downlink SINR in a cellular setup was recently performed in~\cite{7482720,khan2015millimeter}. In~\cite{7482720}, since the exact analysis does not provide insightful results, authors use Frechet's inequality to derive an upper bound on the joint downlink energy and SINR coverage probability. In this paper, we will derive joint energy and downlink SINR coverage probability as the special case of our general result. 

As is evident from the above discussion, all the prior works on stochastic geometry-based analyses of cellular networks with energy harvesting users/devices are either focused on uplink or downlink. To the best of our knowledge, there is no work that deals with joint uplink/downlink coverage probability defined by the joint energy, uplink SINR, and downlink SINR coverage probability, which is the main focus of this paper. That being said, the joint downlink and uplink coverage has received some attention recently in the {\em regularly powered networks}\footnote{Throughput this paper, we will refer to the IoT networks in which the IoT devices have uninterrupted access to a reliable energy source, such as power grid or a battery, as the regularly powered networks.}~\cite{shaikh2016joint,7343565,7112544}. For instance, authors in~\cite{shaikh2016joint} use a 3GPP simulation model to determine whether it is appropriate to assume independence in the uplink and downlink coverage events. The simulation results demonstrate that the two events cannot be treated as independent. This is due to the correlation that results from associating with the same BS in both uplink and downlink. Sometimes this correlation is ignored in the interest of tractability. For instance, in~\cite{7343565}, authors derive the joint uplink/downlink coverage probability as the product of two coverage probabilities. For more accurate analysis, one should of course capture this correlation explicitly, as done in~\cite{7112544}, where the authors provided the accurate joint distribution of uplink and downlink path-loss for generalized uplink/downlink cell association policies (associating with the same BS in both channels is a special case). Assuming independent interference levels over uplink and downlink channels, they use this joint distribution to derive the joint uplink/downlink coverage. 


In this paper we study the performance of {\em on-the-fly} reception/transmission in a cellular-based IoT network where the IoT devices first harvest energy and then use it to receive/transmit information in the same time slot. Assuming cellular transmissions to be the only source of RF energy for the IoT devices, we study the joint probability of a typical IoT device harvesting sufficient energy {\em and} achieving both uplink and downlink SINR thresholds with respect to its associated base station in a given time slot. As noted already, we will refer to this as {\em uplink/downlink coverage probability} in this paper. Since the same infrastructure (cellular BSs) is used for charging and communication, there is inherent correlation in the energy and uplink/downlink coverage events, which is carefully incorporated in our analysis. Please refer to Section~\ref{sec:sys} for more details on the system setup. We now summarize the contributions of this paper.

\subsection{Contributions and Outcomes}
{\em Cellular-based IoT model}. We develop a comprehensive model for cellular-based RF-powered IoT network in which the locations of the BSs and the IoT devices are modeled using two independent Poisson point processes (PPPs). Each time slot is assumed to be partitioned into three sub-slots: (i) {\em charging sub-slot}, in which the received power from the cellular network is used for charging devices to enable them to perform information transmission/reception in the next two sub-slots, (ii) {\em downlink sub-slot}, in which the devices receive information from their associated BSs, and (iii) {\em uplink sub-slot}, in which the devices transmit information to their associated BSs using {\em fractional} channel inversion power control. Contrary to the prior works discussed above that focused on the separate analysis of uplink and downlink coverage, in this paper we focus on the analysis of joint uplink/downlink coverage (defined as the joint probability of energy coverage, uplink SINR coverage, and downlink SINR coverage). Since cellular network is assumed to be the only source of RF energy for the IoT devices, the energy and uplink/downlink coverage events are tightly coupled through the locations of the cellular BSs. In particular, the amount of energy harvested by each device is highly correlated with both the uplink and downlink SINR achieved by that device. Naturally, the uplink and downlink coverage events are also coupled. As discussed next, we carefully handle this correlation in our analysis, which is also one of the main technical contributions of this paper.

{\em Joint uplink/downlink coverage analysis}. As stated already, we define joint uplink/downlink coverage as the joint probability that the typical device harvests sufficient energy in the first sub-slot, achieves high enough downlink SINR in the second sub-slot, and achieves high enough uplink SINR in the third sub-slot. These three events are correlated because of their dependence on the point processes modeling the devices and the base stations. That being said, if we assume independent fading across the three sub-slots and condition on the point processes, the three events become {\em conditionally} independent. We therefore, derive the conditional probabilities of the three events first. The complexity of this problem should be evident from the following two facts: (i) the exact characterization of uplink SINR in a conventional single-tier cellular setup is not known in the stochastic geometry literature~\cite{AndGupJ2016}, and (ii) the total energy harvested is essentially a power-law shot noise field whose probability distribution function is not known in general. On top of these challenges, we need to jointly decondition (average) over the point processes in order to obtain the joint uplink/downlink coverage, which adds to the complexity of the problem. We overcome all these challenges by developing a dominant BS-based approximation approach that not only provides a tight approximation for the power-law shot noise field (energy harvested) but also facilitates joint deconditioning over the point processes. The tightness of the approximate joint coverage expression is verified by comparing it with the simulation results. 

{\em Useful system insights}. Our analytical results provide several useful system insights. First, we demonstrate the existence of optimal time-slot partitioning that maximizes system throughput. The effect of other system parameters on this optimal partitioning is studied numerically. We then compare the performance of the RF-powered IoT system with the one in which IoT devices have access to a reliable power source (termed regularly powered network). Our analytical results reveal several interesting thresholds beyond which the performance of this RF-powered network is similar to that of the regularly powered network. For instance, we show that if the distance of the typical device to the second closest BS is below a certain threshold, its downlink coverage performance would be the same as the regularly powered network. We further study the effect of other system parameters including time-slot partitioning parameters, cellular network density, RF-DC conversion efficiency, and cellular network transmission power on the system performance. We show how these parameters can be tuned in order to get the performance of this RF-powered network closer to that of a regularly powered network. This is done by defining a {\em tuning parameter} that captures the effect of the aforementioned system parameters. Our analysis shows that in order to get the performance of this RF-powered network closer to the regularly powered network, it is only required to make sure that this tuning parameter is large enough. 

\section{System Model}\label{sec:sys}
\begin{figure}
\centering
\includegraphics[width=0.5\columnwidth]{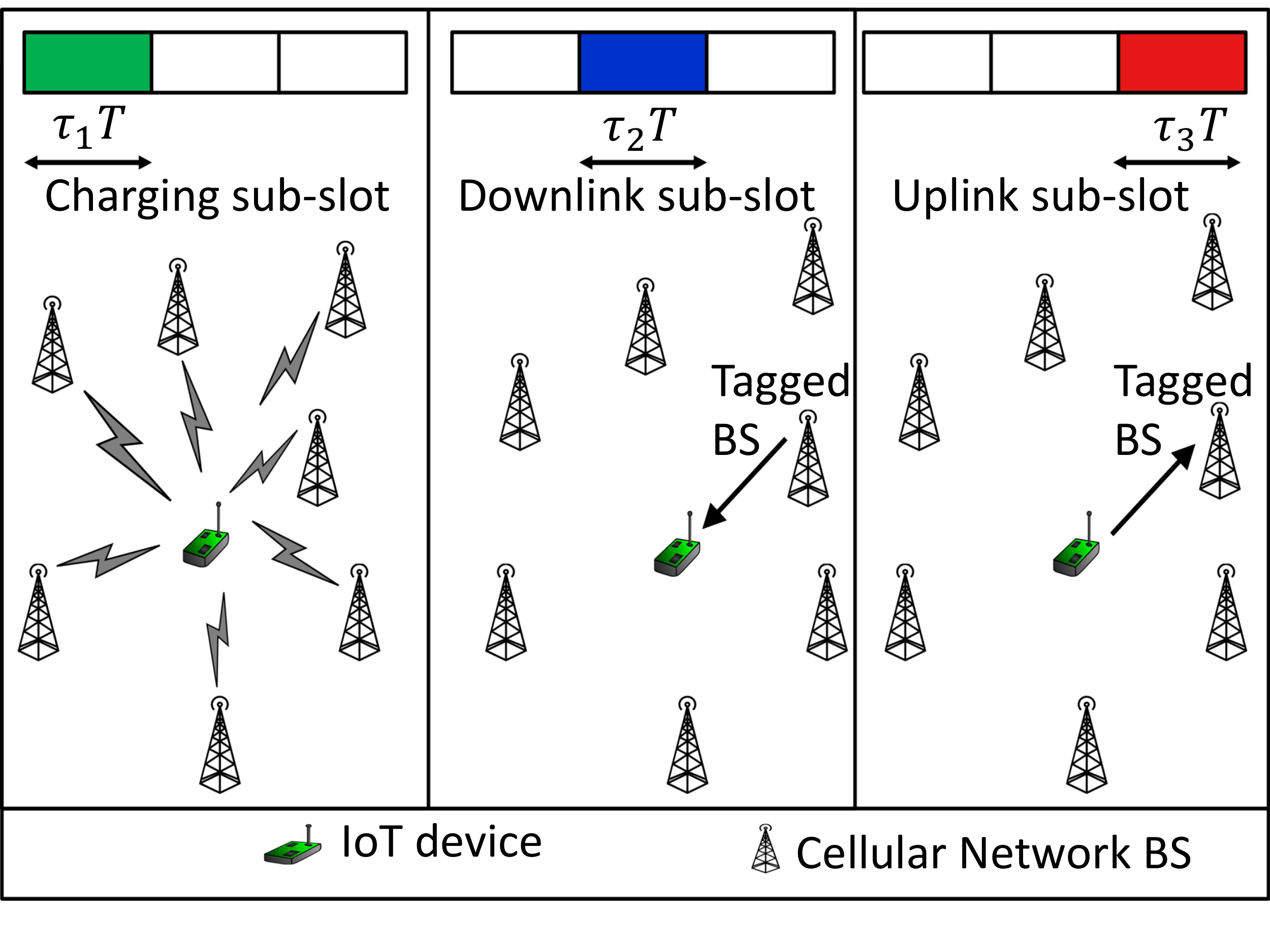}
\caption{Illustration of the system setup and the three sub-slots (charging, downlink, and uplink).}
\label{fig:time_div}
\end{figure}
We consider a cellular-based IoT network in which the IoT devices are solely powered by the ambient RF energy. In this work, we assume that the cellular transmissions are the only source of ambient RF energy for these devices. Quite reasonably, the IoT devices are assumed to be batteryless (similar to~\cite{ozel2011awgn,molavianjazi2015low}). The more general case of finite-sized battery is left for future work. In particular, we assume that all the energy required for uplink and/or downlink communication by a  device in a given time slot will need to be harvested by that device in the same time slot. More details will be provided shortly. The locations of the cellular network BSs and the IoT devices are modeled by two independent PPPs $\Phi_{b}\equiv\{x_{i}\}\subset\mathbb{R}^2$ and $\Phi_u\equiv\{u_{i}\}\subset\mathbb{R}^2$ with densities $\lambda_{b}$ and $\lambda_u$, respectively~\cite{andrews2011tractable}. As will be the case in reality, we assume $\lambda_u > \lambda_{b}$. 

As implied in Fig.~\ref{fig:time_div}, we assume that each IoT device adopts the time-switching receiver architecture (see~\cite{6951347}) in which the antenna is used for energy harvesting for a given fraction of time and for communication for the rest of the time. The time slot duration is assumed to be $T$ (seconds). As shown in Fig.~\ref{fig:time_div}, each time-slot is further divided into charging, downlink, and uplink sub-slots with durations $T_{\rm ch}=\tau_1 T$, $T_{\rm tr}^{\rm DL}=\tau_2 T$, and $T_{\rm tr}^{\rm UL}=\tau_3 T$, respectively. During the {\em charging sub-slot}, all the BSs in the network act as RF chargers for the IoT devices. In the downlink and uplink sub-slots, each IoT device receives and sends information to its associated BS, respectively. This system setup will facilitate the analysis of joint uplink/downlink coverage probability thus generalizing the prior work on energy harvesting networks that focused on the analysis of downlink and uplink separately. Naturally, if we substitute $\tau_2=0$, we can focus only on the uplink analysis, which we refer to as the {\em uplink mode}. Similarly, if we substitute $\tau_3=0$, we can focus only on the downlink analysis, which we refer to as the {\em downlink mode}. The general case in which $\tau_2$ and $\tau_3$ are both non-zero will be referred to as the {\em joint uplink/downlink mode}. Our analysis will be performed under the following assumptions: (i) each IoT device connects to its {\em nearest} BS (referred to as {\em tagged} BS in the rest of the paper), (ii) fading gains across all links are independent, (iii) fading gains across the same link in charging sub-slot (denoted by $g_x$), downlink sub-slot (denoted by $h_x$), and uplink sub-slot (denoted by $w_x$) are independent, (iv) all channels suffer from Rayleigh fading. This means that $g_x$, $h_x$, and $w_x$ are all independent exponential random variables with mean $1$. Under these assumptions, we focus our analysis on a typical device placed at the origin (without loss of generality due to the stationarity of PPP). We now enrich our notation to express key metrics of interest for each sub-slot. 

In the charging sub-slot, we are interested in measuring the amount of energy harvested by the typical device. In order to do that, we first model the received power at the typical device from a BS located at $x \in \Phi_{b}$ as $P_{\rm t}g_{x}\|x\|^{-\alpha}$, where $g_x\sim \exp(1)$ is the fading gain, $P_{\rm t}$ is the transmission power (assumed to be the same for all the BSs), and $\|x\|^{-\alpha}$ models standard power law path-loss with exponent $\alpha>2$. The total energy harvested by the typical device is thus 
\begin{align}
\label{5}
E_{\rm H} =\tau_1 {T}\eta\sum\limits_{x\in \Phi_{b}}P_{\rm t} g_{x}\|x\|^{-\alpha}\ {\rm Joules},
\end{align} 
where $\eta<1$ represents the efficiency of the RF-to-DC conversion. 

In the downlink and uplink sub-slots, we are interested in the expressions for the respective SINRs. For the downlink sub-slot, the the SINR at the typical device is
\begin{equation}
\label{1}
{\rm SINR_{DL}} =\frac{P_{\rm t}h_{x_1}\|x_1\|^{-\alpha}}{\sum_{x\in\Phi_{b}\backslash x_1}P_{\rm t}h_{x}\|x\|^{-\alpha}+\sigma_{\rm DL}^2} = \frac{P_{\rm t}h_{x_1}\|x_1\|^{-\alpha}}{I_1+\sigma_{\rm DL}^2}, 
\end{equation} 
where $h_x\sim\exp(1)$ represents the fading gain between the typical device and the BS located at $x$, $x_1$ is the location of the nearest (tagged) BS, $I_1$ denotes the interference power, and $\sigma_{\rm DL}^2$ models thermal noise power. For successful reception in the downlink sub-slot, the received SINR needs to be greater than a modulation-and-coding specific target SINR $\beta_{\rm DL}$. In addition, the IoT device needs a minimum amount of energy $E_{\rm rec}$ in order to be able to activate its receiving chain circuitry and receive data successfully during the downlink sub-slot. 

\begin{figure}
\centering
\includegraphics[width=0.25\columnwidth]{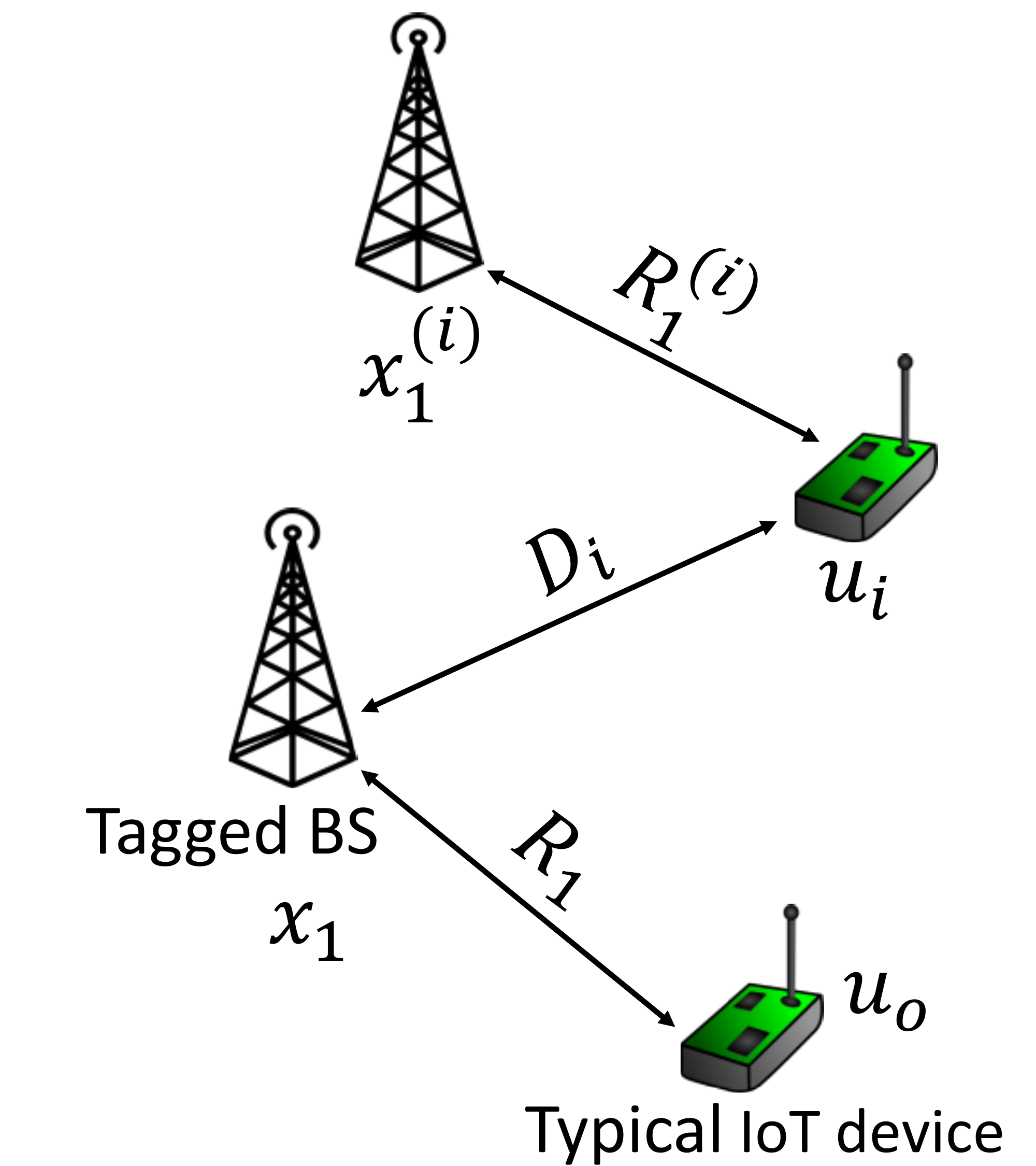}
\caption{Key variables used in the uplink analysis.}
\label{fig:SIR}
\end{figure}
In the uplink sub-slot, each IoT device is assumed to perform uplink fractional channel inversion power control. Hence, if the distance between the IoT device and its serving BS is $R$, then the transmitted power is $\rho R^{\epsilon\alpha}$, where $\rho$ is the BS sensitivity, and $\epsilon \in [0,1]$ is the power control parameter. Therefore, the typical IoT device requires $\tau_3T\rho R^{\epsilon\alpha}$ energy in order to perform uplink transmission. We refer to IoT devices that have enough energy to transmit in the uplink sub-slot as {\em active} devices. Focusing our analysis on the typical device located at the origin, the uplink SINR for this device measured at its tagged BS is: 
\begin{align}
\label{3}
{\rm SINR_{UL}}=\frac{w_o\rho\|x_1\|^{(\epsilon-1)\alpha}}{\sum\limits_{u_i\in\Phi_a\backslash u_o} \delta_iw_i\rho \left(R_1^{(i)}\right)^{\epsilon\alpha}D_i^{-\alpha}+\sigma_{\rm UL}^2},
\end{align}
where $\Phi_a$ is the point process representing all the devices (including the typical device) that are scheduled on the same time-frequency resource as the typical device, $\sigma_{\rm UL}^2$ models thermal noise power, $w_i\sim \exp(1)$ is the channel fading gain between the device located at $u_i$ and the tagged BS during uplink information sub-slot, $x_1$ is the location of the tagged BS, $D_i=\|u_i-x_1\|$ is the distance between the device located at $u_i$ and the tagged BS, $R_1^{(i)}$ is the distance between the device located at $u_i$ and its serving BS (which is the closest BS to this device by definition), and $u_o$ is the location of the typical device. Also $\delta_i$ is an indicator function that equals to $1$ if the IoT device located at $u_i$ is {\em active}, and $0$ otherwise. Please refer to Fig.~\ref{fig:SIR} for the summary of this uplink-specific notation. As was the case in downlink, this SINR needs to be greater than a modulation-and-coding specific target SINR threshold $\beta_{\rm UL}$ for successful transmission.

\begin{table}[t]\caption{Table of Notations}
\centering
\begin{center}
\resizebox{\textwidth}{!}{
\renewcommand{\arraystretch}{1.4}
    \begin{tabular}{ {c} | {c} }
    \hline
        \hline
    \textbf{Notation} & \textbf{Description} \\ \hline
    $\Phi_{b}$; $\lambda_{b}$ & PPP modeling the locations of BSs in the cellular network; density of the BSs \\ \hline
    $\Phi_{u}$; $\lambda_u$ & PPP modeling the locations of IoT devices; density of the IoT devices \\ \hline
        $\tau_1$; $\tau_2$; $\tau_3$ & Time-slot division parameter for charging sub-slot; downlink sub-slot; uplink sub-slot  \\ \hline
        $E_{\rm H}$ & Amount of energy harvested from ambient RF signals (in Joules)\\ \hline
        $D_{\rm avg}$ & Average data rate (throughput)\\ \hline
        $P_{\rm cov}^{\rm DL,RP}$; $P_{\rm suc}^{\rm UL,RP}$ & Downlink coverage probability; uplink coverage probability in a regularly powered network\\ \hline
        $P_{\rm cov}^{\rm DL}$; $P_{\rm suc}^{\rm UL}$; $P_{\rm suc}^{\rm J}$ & Downlink coverage probability; uplink coverage probability; joint uplink/downlink coverage probability in the ambient RF powered network\\ \hline
        $g_x$; $h_x$; $w_x$ & Fading gains during charging; downlink; uplink sub-slots (assumed to be i.i.d. across all links). Rayleigh fading is assumed\\ \hline
        $W_D$ ($W_U)$ & Bandwidth of the downlink channel (uplink channel) \\ \hline
        $P_{\rm t}$; $\rho$ & BS transmission power; BS sensitivity\\ \hline
        $\epsilon$; $\alpha$ & Power control parameter; path loss exponent ($\alpha>2$)\\ \hline
        $r_1$ ($r_2$) & Distance between typical IoT device and its nearest BS (2nd nearest BS)\\ \hline\hline
    \end{tabular}}
\end{center}
\label{tab:TableOfNotations}
\end{table}

{\em Joint uplink/downlink coverage.} With this, we are now ready to formally introduce the key metric of interest for this paper: the joint uplink/downlink coverage probability. 
Recall that the total energy harvested by a device in the charging sub-slot is used by that device to receive information from the tagged BS during the downlink sub-slot and transmit information to the tagged BS during the uplink sub-slot. Hence, the energy coverage condition for this case is: 
\begin{align}
\label{6}
E_{\rm H}\geq E_{\rm min},
\end{align}
where $E_{\rm min}=E_{\rm rec}+\tau_3 T\rho {r_1}^{\epsilon\alpha}$, $r_1=\|x_1\|$ is the distance between the typical IoT device and its nearest BS. For completeness, three conditions need to be satisfied for uplink/downlink coverage: (i) $E_{\rm H}> E_{\rm min}$, (ii) ${\rm SINR_{DL}} > \beta_{\rm DL}$ in the downlink sub-slot, and (iii) ${\rm SINR_{UL}} > \beta_{\rm UL}$ in the uplink sub-slot. Therefore, the  joint uplink/downlink coverage probability is defined as: 
\begin{align}
P_{\rm suc}^{\rm J} & = \mathbb{E}\left[\mathbbm{1}({\rm SINR_{DL}}\geq\beta_{\rm DL})\mathbbm{1}({\rm SINR_{UL}}\geq\beta_{\rm UL})\mathbbm{1}(E_{\rm H}\geq E_{\rm min})\right].
\label{7}
\end{align}
As noted already, when $\tau_2$ and $\tau_3$ are both non-zero, we call this a {\em joint uplink/downlink mode}. As discussed next, if one of them is zero, we can specialize the above definition of joint coverage to study downlink or uplink coverage probability separately. 

{\em Downlink coverage.} If we substitute $\tau_3=0$, each time slot is partitioned into charging and downlink sub-slots. We referred to this as the {\em downlink mode} above. Since each device in this mode only needs to perform downlink transmission, the energy coverage condition reduces to $E_{\rm H}> E_{\rm rec}$. Consequently, the downlink coverage probability for this case can be defined as:
\begin{align}
P_{\rm cov}^{\rm DL} & = \mathbb{E}\left[\mathbbm{1}({\rm SINR_{DL}}\geq\beta_{\rm DL})\mathbbm{1}(E_{\rm H}\geq E_{\rm rec})\right].
\label{2}
\end{align}
Clearly \eqref{2} can be obtained from \eqref{7} by substituting $\tau_3=0$, $\beta_{\rm UL} = 0$ and using $E_{\rm H}> E_{\rm rec}$ as the energy condition. If  ${\rm SINR_{DL}}\geq\beta_{\rm DL}$ and $E_{\rm H}\geq E_{\rm rec}$ (i.e., the IoT device is able to establish a communication link with the BS), the downlink data rate is $R =W_D\log(1+{\beta_{\rm DL}})$ bps in the information sub-slot, where $W_D$ is the bandwidth of the downlink channel.

{\em Uplink coverage.} Similarly, if we substitute $\tau_2=0$, there is no downlink sub-slot and each time slot is partitioned into only charging and uplink sub-slots. This was referred to as the {\em uplink mode} earlier in this Section. Since each IoT device now needs to perform only uplink communication, the energy coverage condition for this case is $E_{\rm H} > \tau_3T\rho R^{\epsilon\alpha}$. This along with the uplink SINR coverage condition gives the following definition for the uplink coverage probability: 
\begin{align}
P_{\rm suc}^{\rm UL} & = \mathbb{E}\left[\mathbbm{1}({\rm SINR_{UL}}\geq\beta_{\rm UL})\mathbbm{1}(E_{\rm H}\geq \tau_3T\rho {r_1}^{\epsilon\alpha})\right].
\label{4}
\end{align}
As was the case for downlink coverage above, \eqref{4} can be obtained from \eqref{7} by substituting $\tau_2=0$, $\beta_{\rm DL} = 0$ and using $E_{\rm H} > \tau_3T\rho r_1^{\epsilon\alpha}$ as the energy condition. If the two coverage conditions (${\rm SINR_{UL}}\geq\beta_{\rm UL}$ and $E_{\rm H}\geq \tau_3T\rho {r_1}^{\epsilon\alpha}$) are satisfied, the uplink data rate is $R=W_U\log(1+{\beta_{\rm UL}})$ bps in the information sub-slot, where $W_U$ is the bandwidth of the uplink channel.

As evident from the above discussion, joint uplink/downlink coverage probability encompasses the other two as special cases. We will therefore start with the analysis of this general case. The results for the downlink and uplink modes will be provided as special cases of this general setup to provide useful system design insights. 

\section{Joint Uplink and Downlink Mode}\label{sec:simult}
This is the first technical section of the paper in which we will evaluate the joint uplink/downlink coverage probability defined in Eq.~\ref{7}. In particular, our goal is to evaluate the joint probability of the following three events: (i) ${\rm SINR_{DL}}\geq\beta_{\rm DL}$, (ii) ${\rm SINR_{UL}}\geq\beta_{\rm UL}$, and (iii) $E_{\rm H}\geq E_{\rm min}$. Keeping the joint treatment aside, the complexity of this analysis should be evident from the following two facts: (i) the exact characterization of $\mathbb{P}\left( {\rm SINR_{UL}}\geq\beta_{\rm UL} \right)$ is not known in the stochastic geometry literature~\cite{AndGupJ2016,HaeJ2017}, and (ii) the total energy harvested is essentially a power-law shot noise field whose probability density function is not known in general. To make matters worse, all these events depend upon the point process $\Phi_b$ modeling the locations of the BSs, which necessitates their joint analysis. This dependence on $\Phi_b$ is quite evident for both $E_{\rm H}$ and ${\rm SINR_{DL}}$ from their expressions given by Eqs. \ref{5} and \ref{1}. While ${\rm SINR_{UL}}$ may not appear to depend on $\Phi_b$ on the first look (see Eq. \ref{3}), the point processes of the devices and BSs are correlated though cell selection and resource scheduling (see \cite{AndGupJ2016} for the detailed discussion), which couples the uplink coverage event with the other two events. Therefore, the main challenge in our analysis is the joint treatment of these three coverage events. That being said, since the main source of this correlation, as evident from Eqs.~\ref{5},~\ref{1},~\ref{3}, is the dependence of the three events on $\Phi_b$, they can be treated as independent when conditioned on $\Phi_{b}$ since the fading gains ($h_x$, $g_x$, and $w_x$) in the three sub-slots are assumed independent. Consequently, the joint uplink/downlink coverage probability defined in Eq.~\ref{7} can be expressed as
\begin{align}
\label{8}
P_{\rm suc}^{\rm J}=\mathbb{E}_{\Phi_{b}}\left[\mathbb{P}\left({\rm SINR_{DL}}\geq\beta_{\rm DL}\Big|\Phi_{b}\right)\mathbb{P}\left({\rm SINR_{UL}}\geq\beta_{\rm UL}\Big|\Phi_{b}\right)\mathbb{P}\left(E_{\rm H}\geq E_{\rm min}\Big|\Phi_{b}\right)\right].
\end{align}
In the following subsections, we carefully approximate the three conditional probability terms using a dominant BS-based approach. The resulting expressions will then be used to derive our main result for the the joint uplink/downlink coverage probability in Theorem~\ref{thm:simult}.

\subsection{Conditional Energy Coverage Probability}

As discussed above, ${\rm SINR_{DL}}$ and $E_{\rm H}$ both depend upon $\Phi_b$ explicitly. However, due to pathloss, the BSs located far away from the typical device do not contribute as much to both these terms as the BSs located close to the typical point. Therefore, we reduce the dimensionality of this problem by considering the effect of closest two BSs to the typical device exactly and approximating the effect of the rest of the BSs. It will be clear shortly why we chose two and not any other number. This dominant BS-based approach is useful when the exact analysis is either too difficult or leads to unwieldy results. It has been used in the past to analyze the coverage of ad hoc networks \cite{WeberJeffJindal}, coverage of downlink cellular networks \cite{6933943}, $k$-coverage of localization networks \cite{SchDhiJ2016,BhaDhiC2016}, and downlink coverage of wireless networks of unmanned aerial vehicles \cite{CheDhiC2016,CheDhiJ2017}. Since all these works focused on some form of (marginal) SINR-based coverage, they are not applicable to our analysis because of the need to perform conditional analysis of each term separately and then decondition jointly over all the terms. These works are listed here mainly for completeness. 

We apply this approach to approximate the total energy harvested by the typical device in the charging sub-slot (given by Eq. \ref{5}) by the energy harvested from the nearest two BSs (located at distances $r_1 = \|x_1\|$ and $r_2 = \|x_2\|$ from the typical device) and the conditional mean (conditioned on the location of the nearest 2 BSs) of the rest of the terms as follows:
\begin{align}
\label{9}
E_{\rm H}=\tau_1 T\eta P_{\rm t}\sum_{x\in \Phi_{b}}g_x\|x\|^{-\alpha}\approx \tau_1 T\eta P_{\rm t} \left(g_{x_1}\|{x_1}\|^{-\alpha}+g_{x_2}\|{x_2}\|^{-\alpha}+\Psi(r_2)\right),
\end{align}
where $\Psi(r_2)= \mathbb{E}\left[\sum_{x\in \Phi_{b}\backslash{x_1,x_2}}g_x\|x\|^{-\alpha}\Big|x_1,x_2\right]$. We will use this approximation to compute the conditioned energy coverage probability $\mathbb{P}(E_{H}\geq\E_{\rm min}|\Phi_{b})$ which is necessary for the computation of $P_{\rm suc}^{\rm J}$ as explained above. In addition to enabling the joint coverage analysis, this approximation will also lead to several crisp system design insights. For instance, as a result of using this approximation, we will be able to define a threshold on $r_2$ (as well as $\lambda_b$ and time switching parameters) below which the performance is approximately equivalent to that of a regularly powered cellular-based IoT (further discussion will be provided in Remarks~\ref{rem:Pcov2} and~\ref{rem:Psuc3}). As discussed already, the typical IoT device needs to harvest a minimum amount of energy $E_{\rm min}=E_{\rm rec}+\tau_3 T\rho\|x_1\|^{\epsilon\alpha}$ to be able to receive and transmit information. If it is able to harvest this energy, it is said to be in {\em energy coverage}. In the following Lemma we derive an expression for the conditional energy coverage probability using the approximation in Eq.~\ref{9}.

\begin{lemma}[Conditional Energy Coverage Probability]\label{lem:energy_cov_simult}
Probability that the harvested energy during the charging sub-slot is greater than $E_{\rm min}$ conditioned on the point process $\Phi_{b}$ is 
\begin{align}
\label{10}
\mathbb{P}\left(E_{\rm H}\geq E_{\rm min}\Big|\Phi_{b}\right)&=
\frac{r_2^{\alpha}\exp\left(-r_1^{\alpha}\left[\mathcal{F}(r_1,r_2)\right]^{+}\right)-r_1^{\alpha}\exp\left(-r_2^{\alpha}\left[\mathcal{F}(r_1,r_2)\right]^{+}\right)}{r_2^{\alpha}-r_1^{\alpha}},
\end{align}
while the unconditioned probability is
\begin{align}
\label{11}
\mathbb{P}\left(E_{\rm H}\geq E_{\rm min}\right)&=
\int\limits_{0}^{\infty}\int\limits_{r_1\in\mathcal{N}_{r_2}} f_{R_1,R_2}(r_1,r_2){\rm d}r_1{\rm d}r_2\\
&+\int\limits_{0}^{\infty}\int\limits_{r_1\in\mathcal{P}_{r_2}} f_{R_1,R_2}(r_1,r_2)\frac{r_2^{\alpha}\exp\left(-r_1^{\alpha}\mathcal{F}(r_1,r_2)\right)-r_1^{\alpha}\exp\left(-r_2^{\alpha}\mathcal{F}(r_1,r_2)\right)}{r_2^{\alpha}-r_1^{\alpha}}{\rm d}r_1{\rm d}r_2,\nonumber
\end{align} 
where $\mathcal{F}(r_1,r_2)=\left[C(\tau_1)+\frac{\tau_3 \rho r_1^{\epsilon\alpha}}{\tau_1 \eta P_{\rm t}}-\frac{2\pi\lambda_{b}}{\alpha-2} r_2^{2-\alpha}\right]$, $C(\tau_1)=\frac{E_{\rm rec}}{\tau_1 T\eta P_{\rm t}}$, $[x]^{+}=\max\{0,x\}$, $f_{R_1,R_2}(r_1,r_2)=(2\pi\lambda_{b})^2 r_1r_2e^{-\lambda_{b}\pi {r_2}^2}$, $\mathcal{N}_{r_2}=\{r_1:\mathcal{F}(r_1,r_2)\leq 0,\ r_1<r_2\}$, and $\mathcal{P}_{r_2}=\{r_1:\mathcal{F}(r_1,r_2)\geq 0,\ r_1<r_2\}$. 
\end{lemma}
\begin{IEEEproof}
See Appendix~\ref{app:energy_cov_simult}.
\end{IEEEproof}
As explained before, the above expression can be used to compute the energy coverage probability in the downlink mode by eliminating uplink conditions and vice versa for the uplink mode. The complete results for these special cases will be presented in Lemmas~\ref{lem:energy_cov} and~\ref{lem:energy_cov_UL}. 

\subsection{Conditional SINR Coverage Probability} 
As a result of using the approximation introduced in Eq.~\ref{9}, the conditional energy coverage probability in Eq.~\ref{10} is only a function of the distances $r_1$ and $r_2$ (between the typical device and its nearest two BSs). Keeping in mind that we will have to jointly decondition on all the coverage events at the end (as evident from Eq.~\ref{8}), it will be useful to derive conditional downlink SINR coverage also in terms of $r_1$ and $r_2$. In order to do that, we use the same dominant BS-based approach that we used in the previous Subsection. In particular, we approximate the interference in the denominator of SINR in Eq.~\ref{1} by the interference from the second nearest BS (strongest interferer) and the expectation of the interference from the rest of the BSs. Under this approximation, the conditional downlink SINR coverage probability becomes $\mathbb{P}\left({\rm SINR_{DL}}\geq\beta_{\rm DL}\Big|r_1,r_2\right)$. A tractable expression for this conditional probability is derived next. 

\begin{lemma}[Conditional Downlink SINR Coverage Probability]\label{lem:DL_SINR_simult}
Probability that the downlink SINR at the typical device exceeds $\beta_{\rm DL}$, conditioned on $r_1$ and $r_2$, is 
\begin{align}
\label{12}
\mathbb{P}\left({\rm SINR_{DL}}\geq\beta_{\rm DL}\Big|r_1,r_2\right)&= \exp\left(-\mathcal{G}(r_1,r_2)\right)\frac{1}{1+\frac{\beta_{\rm DL}r_1^{\alpha}}{r_2^{\alpha}}},
\end{align}
\end{lemma}
where $\mathcal{G}(r_1,r_2)=\frac{\beta_{\rm DL}\sigma_{\rm DL}^2r_1^{\alpha}}{P_{\rm t}}+\frac{2\pi\lambda\beta_{\rm DL}r_1^{\alpha}}{(\alpha-2)r_2^{\alpha-2}}$.
\begin{IEEEproof}
See Appendix~\ref{app:SINR_simult}.
\end{IEEEproof}
With this, we are now left with deriving the conditional uplink probability, which we do next. It is noteworthy that uplink analysis is known to be a challenging problem even for regularly powered networks. The locations of the devices scheduled in the same time frequency resource block as the typical device (modeled as point process $\Phi_a\backslash u_o$ in Eq.~\ref{3}) are correlated with the locations of the BSs due to the structure of the Poisson Voronoi tessellation. This correlation is further enhanced due to uplink power control, where the transmission power of each device is a function of its distance to its serving BS. As discussed in \cite{AndGupJ2016}, the exact analysis of this setup in not known. It has, however, been shown that modeling the locations of the devices by an independent PPP and handling dependence between the distances $D_i$ and $R_1^{(i)}$ (as defined in Eq.~\ref{3}) appropriately leads to a fairly tight approximation. For the latter, it is sufficient to just account for the fact that $R_1^{(i)} < D_i$, i.e., the serving BS must be closer to the interfering device than the tagged BS. Please refer to \cite{AndGupJ2016} for more details. 
Using this general idea, the Laplace transform of the aggregate interference $I_2={\sum\limits_{u_i\in\Phi_a\backslash u_o} w_i\left(R_1^{(i)}\right)^{\epsilon\alpha}D_i^{-\alpha}}$ at the tagged BS in a regularly powered network was given in~\cite{AndGupJ2016} as follows:  
\begin{align}
\label{13}
&\mathcal{L}_{I_2}(s)=\mathbb{E}\left[e^{-I_2s}\right]=\exp\left(-2\pi{\lambda_b}\int_0^\infty\int_0^{x^2}\frac{1}{1+(s)^{-1}u^{-\alpha\epsilon/2}x^{\alpha}}\pi{\lambda_b}e^{-{\lambda_b}\pi u}{\rm d}ux{\rm d}x\right),
\end{align}
where $\Phi_a$ is the point process modeling the locations of the selected devices in a given time-frequency resource. This expression was used to derive the uplink coverage probability for regularly powered networks in~\cite{AndGupJ2016} as follows:
\begin{align}
\label{14}
P_{\rm suc}^{\rm UL,RP}&=\int_0^\infty f_{R_1}(r_1) e^{\left(-\frac{\beta_{\rm UL}\sigma_{\rm UL}^2}{\rho {r_1}^{(\epsilon-1)\alpha}}\right)}\mathcal{L}_{I_2}\left(\frac{\beta_{\rm UL}}{{r_1}^{(\epsilon-1)\alpha}}\right) {\rm d}r_1,
\end{align}
where $f_{R_1}(r_1)=2\pi\lambda_br_1\exp(-\pi\lambda_br_1^2)$. We will use this expression to compare the performance of the proposed setup to that of the regularly powered networks. 

Coming to the conditional uplink coverage in the proposed energy harvesting setup, note that the dominant source of correlation between uplink SINR and the other two  terms (downlink SINR and the amount of energy harvested) is the serving distance $r_1$. If we condition on $r_1$ and treat $\Phi_a$ and $\Phi_b$ as independent point processes (as done above), the conditional uplink coverage probability reduces to $\mathbb{P}\left({\rm SINR_{UL}}\geq\beta_{\rm UL}\Big|r_1\right)$, which is derived in the next Lemma. 


\begin{lemma}[Conditional Uplink SINR Coverage Probability]\label{lem:UL_SINR_simult}
Probability that the uplink SINR of the typical device at the tagged BS is greater than $\beta_{\rm UL}$, conditioned on $r_1$, is 
\begin{align}
\label{15}
\mathbb{P}\left({\rm SINR_{UL}}\geq\beta_{\rm UL}\Big|r_1\right)&= e^{\left(-\frac{\beta_{\rm UL}\sigma_{\rm UL}^2}{\rho {r_1}^{(\epsilon-1)\alpha}}\right)}\mathcal{L}_{\tilde{I}_2}\left(\frac{\beta_{\rm UL}}{{r_1}^{(\epsilon-1)\alpha}}\right),
\end{align}
where $\mathcal{L}_{\tilde{I}_2}(s)$ is given by Eq.~\ref{13} by replacing $\lambda_b$ with $\tilde{\lambda}_b=P_h \lambda_b$, where $P_h=\mathbb{P}(E_{\rm H}\geq E_{\rm min})$.
\end{lemma}
\begin{IEEEproof}
See Appendix~\ref{app:SINR_simult}.
\end{IEEEproof}

\subsection{Joint Uplink/Downlink Coverage Probability}
Having derived the three conditional probability terms appearing in Eq.~\ref{8} in Lemmas~\ref{lem:energy_cov_simult},~\ref{lem:DL_SINR_simult}, and~\ref{lem:UL_SINR_simult}, we are now ready to derive the joint uplink/downlink coverage probability. The only remaining step is to uncondition their product with respect to the joint distribution of $r_1$ and $r_2$, which results in the following Theorem.
\begin{theorem}[Joint uplink/downlink coverage probability] \label{thm:simult}
The joint uplink/ downlink coverage probability $P_{\rm suc}^{\rm J}$ of the typical IoT device with downlink and uplink SINR thresholds $\beta_{\rm DL}$ and $\beta_{\rm UL}$ respectively is given by:
\begin{align}
\label{16}
&P_{\rm suc}^{\rm J}=\int\limits_0^{\infty}\int\limits_{r_1\in\mathcal{N}_{r_2}} f_{R_1,R_2}(r_1,r_2) e^{\left(-\frac{\beta_{\rm UL}\sigma_{\rm UL}^2}{\rho {r_1}^{(\epsilon-1)\alpha}}\right)}\mathcal{L}_{\tilde{I}_2}\left(\frac{\beta_{\rm UL}}{{r_1}^{(\epsilon-1)\alpha}}\right)\exp\left(-\mathcal{G}(r_1,r_2)\right)\frac{1}{1+\frac{\beta_{\rm DL}r_1^{\alpha}}{r_2^{\alpha}}} {\rm d}r_1{\rm d}r_2\nonumber\\
&+\int\limits_0^{\infty}\int\limits_{r_1\in\mathcal{P}_{r_2}} f_{R_1,R_2}(r_1,r_2) e^{\left(-\frac{\beta_{\rm UL}\sigma_{\rm UL}^2}{\rho {r_1}^{(\epsilon-1)\alpha}}\right)}\mathcal{L}_{\tilde{I}_2}\left(\frac{\beta_{\rm UL}}{{r_1}^{(\epsilon-1)\alpha}}\right)\frac{\exp\left(-\mathcal{G}(r_1,r_2)-r_1^{\alpha}\mathcal{F}(r_1,r_2)\right)}{r_2^{\alpha}-r_1^{\alpha}}\frac{r_2^{\alpha}}{1+\frac{\beta_{\rm DL}r_1^{\alpha}}{r_2^{\alpha}}}{\rm d}r_1{\rm d}r_2\nonumber\\
&-\int\limits_0^{\infty}\int\limits_{r_1\in\mathcal{P}_{r_2}} f_{R_1,R_2}(r_1,r_2) e^{\left(-\frac{\beta_{\rm UL}\sigma_{\rm UL}^2}{\rho {r_1}^{(\epsilon-1)\alpha}}\right)}\mathcal{L}_{\tilde{I}_2}\left(\frac{\beta_{\rm UL}}{{r_1}^{(\epsilon-1)\alpha}}\right)\frac{\exp\left(-\mathcal{G}(r_1,r_2)-r_2^{\alpha}\mathcal{F}(r_1,r_2)\right)}{r_2^{\alpha}-r_1^{\alpha}}\frac{r_1^{\alpha}}{1+\frac{\beta_{\rm DL}r_1^{\alpha}}{r_2^{\alpha}}}{\rm d}r_1{\rm d}r_2,
\end{align} 
where $f_{R_1,R_2}(r_1,r_2)=(2\pi\lambda_b)^2r_1r_2e^{-\pi\lambda_b r_2^2}$, $\mathcal{F}(r_1,r_2)$, $\mathcal{N}_{r_2}$, $\mathcal{P}_{r_2}$ are as introduced in Lemma~\ref{lem:energy_cov_simult}, $\mathcal{G}(r_1,r_2)$ is as introduced in Lemma~\ref{lem:DL_SINR_simult}, and $\mathcal{L}_{\tilde{I}_2}(s)$ is as introduced in Lemma~\ref{lem:UL_SINR_simult}. 
\end{theorem}
\begin{IEEEproof}
This result follows directly by substituting Eq.~\ref{10},~\ref{12},~\ref{15} in Eq.~\ref{8} and integrating over $r_1$ and $r_2$ using the joint distribution $f_{R_1,R_2}(r_1,r_2)$ as defined in~\cite[Eq.~28]{moltchanov2012distance}.
\end{IEEEproof}

\begin{remark} \label{rem:simult}
This general result can be used to derive both downlink coverage and uplink coverage probabilities defined in Eqs.~\ref{2} and~\ref{4}. For instance, if we remove all the uplink conditions by putting $\beta_{\rm UL}=0$ (note that $\mathcal{L}_{I_2}(0)=1$) and $\tau_3=0$, then Eq.~\ref{16} will represent the downlink coverage probability $P_{\rm cov}^{DL}$ for the downlink mode. Similarly, if we remove all the downlink conditions by putting $\beta_{\rm DL}=0$ (note that $\mathcal{G}(r_1,r_2)=0$ in that case), $E_{\rm rec}=0$, and $\tau_2=0$, then Eq.~\ref{16} will represent the uplink coverage probability $P_{\rm suc}^{\rm UL}$ for the uplink mode.%
\end{remark}

\subsection{Average Throughput}
We now derive expressions for both the uplink and the downlink average throughput in the joint uplink/downlink mode. The average downlink throughput is
\begin{align}
\label{17}
D_{\rm avg}^{\rm DL}&=\tau_2 R_{\rm avg}^{\rm DL}=\tau_2 \mathbb{E}\left[W_D\log(1+{\beta_{\rm DL}})\mathbbm{1}({\rm SINR_{DL}}\geq\beta_{\rm DL})\mathbbm{1}({\rm SINR_{UL}}\geq\beta_{\rm UL})\mathbbm{1}(E_{\rm H}\geq E_{\rm min})\right]\nonumber\\
&=\tau_2 W_D\log(1+{\beta_{\rm DL}})P_{\rm suc}^{\rm J},
\end{align}
where $R_{\rm avg}^{\rm DL}$ is the average data rate during downlink sub-slot in the joint mode. The multiplication by $\tau_2$ accounts for the fact that downlink sub-slot lasts for $\tau_2$ fraction of the total time-slot duration. Similarly, the average uplink throughput in the joint mode is:
\begin{align}
\label{18}
D_{\rm avg}^{\rm UL}&=\tau_3 R_{\rm avg}^{\rm UL}=\tau_3 \mathbb{E}\left[W_U\log(1+{\beta_{\rm UL}})\mathbbm{1}({\rm SINR_{UL}}\geq\beta_{\rm UL})\mathbbm{1}({\rm SINR_{DL}}\geq\beta_{\rm DL})\mathbbm{1}(E_{\rm H}\geq E_{\rm min})\right]\nonumber\\
&=\tau_3 W_U\log(1+{\beta_{\rm UL}})P_{\rm suc}^{\rm J},
\end{align}
where $R_{\rm avg}^{\rm UL}$ is the average data rate during uplink sub-slot in the joint mode. 

\begin{remark} \label{rem:Ravg_simult}
Note that for a given $\tau_3$, it is easier to satisfy the energy constraint for larger values of $\tau_1$. This means both $P_{\rm suc}^{\rm J}$ and $R_{\rm avg}^{\rm DL}$ are the increasing functions of $\tau_1$. However, increasing $\tau_1$ decreases $\tau_2$ (for a given $\tau_3$), which reduces the downlink transmission time and may therefore reduce average data rate $D_{\rm avg}^{\rm DL}$. This indicates the existence of an optimal slot partitioning for maximizing $D_{\rm avg}^{\rm DL}$. Similar conclusions can be drawn about the relation between $\tau_1$ and $D_{\rm avg}^{\rm UL}$ for a given $\tau_2$. We will discuss more about this optimal slot partitioning in the sequel.
\end{remark}

In the next two Sections, we will specialize the general results of this Section to the downlink and uplink modes, which will provide several useful system design insights. 

\section{Downlink Mode}\label{sec:DL}
The coverage probability in the downlink mode defined in Eq.~\ref{2} can be expressed as
\begin{align}
\label{19}
P_{\rm cov}^{\rm DL}=\mathbb{E}_{\Phi_{b}}\left[\mathbb{P}\left({\rm SINR_{DL}}\geq\beta_{\rm DL}\Big|\Phi_{b}\right)\mathbb{P}\left(E_{\rm H}\geq E_{\rm rec}\Big|\Phi_{b}\right)\right],
\end{align}
which is the special case of Eq.~\ref{8}. As discussed in Section~\ref{sec:sys} and later in Remark~\ref{rem:simult}, we can obtain this definition for $P_{\rm cov}^{\rm DL}$ by simply substituting $\tau_3=0$, $\beta_{\rm UL} = 0$ and using $E_{\rm H}> E_{\rm rec}$ as the energy condition in the definition of joint uplink/downlink coverage probability given in Eq.~\ref{7} (and hence Eq.~\ref{8}). Therefore, $P_{\rm cov}^{\rm DL}$ can be derived directly by making these substitutions in Eq.~\ref{16}. Similarly, we can derive the energy coverage for downlink mode by using the same substitutions in Eq.~\ref{11}. We first state this energy coverage result next. 
\begin{lemma}[Energy coverage probability in the downlink mode]\label{lem:energy_cov}
Probability that the harvested energy during the charging sub-slot is greater than the value $E_{\rm rec}$ is
\begin{align}
\label{20}
\mathbb{P}(E_{\rm H}\geq E_{\rm rec}) &=1-\exp\left(-\pi\lambda_b\mathcal{A}^2\right)-\pi\lambda_b\mathcal{A}^2\exp\left(-\pi\lambda_b\mathcal{A}^2\right)\nonumber\\
&+\int\limits_{\mathcal{A}}^{\infty}\int\limits_{0}^{r_2} f_{R_1,R_2}(r_1,r_2)\frac{r_2^{\alpha}\exp\left(-r_1^{\alpha}\mathcal{F}_{\rm DL}(r_1,r_2)\right)-r_1^{\alpha}\exp\left(-r_2^{\alpha}\mathcal{F}_{\rm DL}(r_1,r_2)\right)}{r_2^{\alpha}-r_1^{\alpha}}{\rm d}r_1{\rm d}r_2,
\end{align}
where $f_{R_1,R_2}(r_1,r_2)=(2\pi\lambda_b)^2r_1r_2e^{-\pi\lambda_br_2^2}$, $\mathcal{F}_{\rm DL}(r_1,r_2) = C(\tau_1)-\frac{2\pi\lambda_br_2^{2-\alpha}}{\alpha-2}$, $C(\tau_1)=\frac{E_{\rm rec}}{\tau_1 T\eta P_{\rm t}}$, and $\mathcal{A}=\left(\frac{2\pi\lambda_{b}}{C(\tau_1)(\alpha-2)}\right)^\frac{1}{\alpha-2}$. 
\end{lemma}
\begin{IEEEproof}
See Appendix~\ref{app:DL}.
\end{IEEEproof}
\begin{remark} \label{rem:energy_cov}
The effect of the duration of the charging sub-slot $T_{\rm ch}=\tau_1 T$ appears in the value of $C(\tau_1)$. Consistent with intuition, as this duration increases, the value of $C(\tau_1)$ decreases and the energy coverage probability approaches $1$.
\end{remark}
We now state the (downlink) coverage result for the downlink mode (defined in Eq.~\ref{2}). 
\begin{theorem}[Downlink coverage probability in the downlink mode] \label{thm:Pcov}
The downlink coverage probability with SINR threshold $\beta_{\rm DL}$ and minimum required energy $E_{\rm rec}$ is given by
\begin{align}
\label{21}
&P_{\rm cov}^{\rm DL}= \int\limits_0^{\mathcal{A}}\int\limits_{0}^{r_2} f_{R_1,R_2}(r_1,r_2)\exp(-\mathcal{G}(r_1,r_2))\frac{1}{1+\frac{\beta_{\rm DL}r_1^{\alpha}}{r_2^{\alpha}}}{\rm d}r_1{\rm d}r_2\\
&+\int\limits_{\mathcal{A}}^{\infty}\int\limits_{0}^{r_2} f_{R_1,R_2}(r_1,r_2)\exp(-\mathcal{G}(r_1,r_2))\frac{r_2^{\alpha}\exp\left(-r_1^{\alpha}\mathcal{F}_{\rm DL}(r_1,r_2)\right)-r_1^{\alpha}\exp\left(-r_2^{\alpha}\mathcal{F}_{\rm DL}(r_1,r_2)\right)}{\left(r_2^{\alpha}-r_1^{\alpha}\right)\left(1+\frac{\beta_{\rm DL}r_1^{\alpha}}{r_2^{\alpha}}\right)}{\rm d}r_1{\rm d}r_2,\nonumber
\end{align}
where $\mathcal{G}(r_1,r_2)$ is defined in Lemma~\ref{lem:DL_SINR_simult}, $C(\tau_1)$, $\mathcal{A}$, and $\mathcal{F}_{\rm DL}(r_1,r_2)$ are defined in Lemma~\ref{lem:energy_cov}.
\end{theorem}
\begin{IEEEproof}
See Appendix~\ref{app:DL}.
\end{IEEEproof}
\begin{remark} \label{rem:Pcov}
The effect of the duration of the charging sub-slot $T_{\rm ch}=\tau_1 T$ appears mainly in the value of $\mathcal{A}$ (implicitly in the value of $C(\tau_1)$). It can be observed that as this duration increases, the value of $\mathcal{A}$ increases and $P_{\rm cov}^{\rm DL}$ approaches the coverage probability of regularly powered network $P_{\rm cov}^{\rm DL,RP}=\mathbb{P}({\rm SINR}\geq \beta_{\rm DL})$. This is because as $T_{\rm ch}$ increases, it  becomes easier to satisfy the energy constraint and the energy coverage probability ultimately approaches unity.
\end{remark}
\begin{remark} \label{rem:Pcov2}
Conditioned on $\Phi_b$, the variable $\mathcal{A}$ represents an important system parameter. In Eq.~\ref{21}, it can be interpreted as a threshold on the value of $r_2$. In particular, as long as the distance to the second nearest BS (which is also the second dominant RF source on average) is less than this threshold, the energy coverage condition is satisfied and the only condition required for coverage is ${\rm SINR_{DL}}\geq\beta_{\rm DL}$, which is represented by the first term in Eq.~\ref{21}. This useful system insight is a result of using the approximation in Eq.~\ref{9} that defines the amount of energy harvested in terms of distances $r_1$ and $r_2$. This provides useful characterization of the regime in which the performance of this RF-powered IoT network will be similar to the regularly powered network. Similar observations will be provided for the uplink case in the next Section.%
\end{remark}
The general expression for average throughput given in Eq.~\ref{17} can be specialized for the downlink mode as follows: 
\begin{align}
\label{22}
D_{\rm avg}^{\rm DL}&= \tau_2W_D\log(1+{\beta_{\rm DL}})P_{\rm cov}^{\rm DL}.
\end{align}
\begin{remark} \label{rem:Ravg}
Similar to our comments in Remark~\ref{rem:Ravg_simult}, $R_{\rm avg}^{\rm DL}$ is an increasing function of $\tau_1$. On the other hand, the duration $\tau_2T$ of the downlink sub-slot  decreases with increase in $\tau_1$. This indicates the existence of an optimal value of $\tau_1$ that maximizes $D_{\rm avg}^{\rm DL}$. 
\end{remark}

\section{Uplink Mode}\label{sec:UL}
In this Section, we specialize the results of Section~\ref{sec:simult} to the uplink mode. Recall that in the uplink mode, each time-slot is partitioning into two sub-slots: charging sub-slot and uplink sub-slot. As discussed in Section \ref{sec:sys} and Remark \ref{rem:simult}, the uplink coverage probability, defined in Eq. \ref{4}, can be obtained from the definition of joint uplink/downlink coverage, given by Eq. \ref{7}, by substituting $\tau_2=0$, $\beta_{\rm DL} = 0$ and using $E_{\rm H} > \tau_3T\rho r_1^{\epsilon\alpha}$ as the energy condition. Consequently, the results for the uplink mode can be obtained from the general results of Section~\ref{sec:simult} by making these substitutions. While these substitutions are quite similar to the ones that we made in the previous Section for the downlink mode, there is a subtle difference in the energy conditions, which is the reason why the final results are slightly different in the two cases. In particular, while the minimum required energy in the downlink mode was fixed ($E_{\rm rec}$), it is a function of the nearest BS location in the uplink mode (due to power control). As in the previous Section, we first state the energy coverage result for the Uplink mode next. 

\begin{lemma}[Energy coverage probability in the uplink mode]\label{lem:energy_cov_UL}
Energy coverage probability is 
\begin{align}
\label{23}
\mathbb{P}&(E_{\rm H}\geq \tau_3T\rho\|x_1\|^{\epsilon\alpha})=1-\exp\left(-\pi\lambda_b\tilde{\mathcal{A}}^2\right)-\pi\lambda_b\tilde{\mathcal{A}}^2\exp\left(-\pi\lambda_b\tilde{\mathcal{A}}^2\right)+\int\limits_{\tilde{\mathcal{A}}}^{\infty}\int\limits_0^{\mathcal{H}(r_2)} f_{R_1,R_2}(r_1,r_2){\rm d}r_1{\rm d}r_2\nonumber\\
&+\int\limits_{\tilde{\mathcal{A}}}^{\infty}\int\limits_{\mathcal{H}(r_2)}^{r_2} f_{R_1,R_2}(r_1,r_2)\frac{r_2^{\alpha}\exp\left(-r_1^{\alpha}\mathcal{F}_{\rm UL}(r_1,r_2)\right)-r_1^{\alpha}\exp\left(-r_2^{\alpha}\mathcal{F}_{\rm UL}(r_1,r_2)\right)}{r_2^{\alpha}-r_1^{\alpha}}{\rm d}r_1{\rm d}r_2,
\end{align}
where $f_{R_1,R_2}(r_1,r_2)=(2\pi\lambda_b)^2r_1r_2e^{-\pi\lambda_br_2^2}$, $\mathcal{H}(r_2)=\left(\frac{2\pi\lambda_b}{(\alpha-2)\tilde{C}(\tau_1)}\right)^{\frac{1}{\epsilon\alpha}}r_2^{\frac{2-\alpha}{\epsilon\alpha}}$, $\mathcal{F}_{\rm UL}(r_1,r_2) = \tilde{C}(\tau_1)r_1^{\epsilon\alpha}-\frac{2\pi\lambda_br_2^{2-\alpha}}{\alpha-2}$, $\tilde{C}(\tau_1)=\frac{\tau_3\rho}{\tau_1\eta P_t}$, $\tilde{\mathcal{A}}=\left(\frac{2\pi\lambda_{b}}{\tilde{C}(\tau_1)(\alpha-2)}\right)^\frac{1}{(\epsilon+1)\alpha-2}$. 
\end{lemma}
\begin{IEEEproof}
See Appendix~\ref{app:UL}.
\end{IEEEproof} 
\begin{remark} \label{rem:energy_cov_UL}
It is easy to see that increasing the density $\lambda_{b}$ of the BS PPP $\Phi_{b}$ increases energy coverage probability due to two reasons. First, it reduces the distance $r_1$ between the typical device and its serving BS, which reduces the transmission power ${r_1}^{\epsilon\alpha}$ of this device, this making it easier to satisfy the energy coverage condition. Second, increasing $\lambda_{b}$ also increases the aggregate energy $E_{\rm H}$ harvested by the typical device. This is also evident from Eq.~\ref{23} where all the terms can be shown to be decreasing functions of $\lambda_b$. 
\end{remark} 
We now present the uplink coverage probability (defined in Eq.~\ref{4}) next. Using this, we will discuss the differences between the regularly powered and energy harvesting networks.
\begin{theorem}[Uplink coverage probability in the uplink mode] \label{thm:Psuc}
The uplink coverage probability $P_{\rm suc}^{\rm UL}$ of the IoT device with SINR threshold $\beta_{\rm UL}$ and uplink transmission power $\rho\|x_1\|^{\epsilon\alpha}$ is 
\begin{align}
\label{24}
P_{\rm suc}^{\rm UL}&=\int\limits_0^{\tilde{\mathcal{A}}}\int\limits_0^{r_2} f_{R_1,R_2}(r_1,r_2)e^{\left(-\frac{\beta_{\rm UL}\sigma_{\rm UL}^2}{\rho {r_1}^{(\epsilon-1)\alpha}}\right)}\mathcal{L}_{\tilde{I}_2}\left(\frac{\beta_{\rm UL}}{{r_1}^{(\epsilon-1)\alpha}}\right){\rm d}r_1{\rm d}r_2\nonumber\\
&+\int\limits_{\tilde{\mathcal{A}}}^{\infty}\int\limits_0^{\mathcal{H}(r_2)} f_{R_1,R_2}(r_1,r_2)e^{\left(-\frac{\beta_{\rm UL}\sigma_{\rm UL}^2}{\rho {r_1}^{(\epsilon-1)\alpha}}\right)}\mathcal{L}_{\tilde{I}_2}\left(\frac{\beta_{\rm UL}}{{r_1}^{(\epsilon-1)\alpha}}\right){\rm d}r_1{\rm d}r_2\nonumber\\
&+\int\limits_{\tilde{\mathcal{A}}}^{\infty}\int\limits_{\mathcal{H}(r_2)}^{r_2} f_{R_1,R_2}(r_1,r_2)\frac{r_2^{\alpha}\exp\left(-r_1^{\alpha}\mathcal{F}_{\rm UL}(r_1,r_2)-\frac{\beta_{\rm UL}\sigma_{\rm UL}^2}{\rho {r_1}^{(\epsilon-1)\alpha}}\right)}{r_2^{\alpha}-r_1^{\alpha}}\mathcal{L}_{\tilde{I}_2}\left(\frac{\beta_{\rm UL}}{{r_1}^{(\epsilon-1)\alpha}}\right){\rm d}r_1{\rm d}r_2\nonumber\\
&-\int\limits_{\tilde{\mathcal{A}}}^{\infty}\int\limits_{\mathcal{H}(r_2)}^{r_2} f_{R_1,R_2}(r_1,r_2)\frac{r_1^{\alpha}\exp\left(-r_2^{\alpha}\mathcal{F}_{\rm UL}(r_1,r_2)-\frac{\beta_{\rm UL}\sigma_{\rm UL}^2}{\rho {r_1}^{(\epsilon-1)\alpha}}\right)}{r_2^{\alpha}-r_1^{\alpha}}\mathcal{L}_{\tilde{I}_2}\left(\frac{\beta_{\rm UL}}{{r_1}^{(\epsilon-1)\alpha}}\right){\rm d}r_1{\rm d}r_2,
\end{align} 
where $\mathcal{H}(r_2)$, $\tilde{C}(\tau_1)$, $\mathcal{F}_{\rm UL}(r_1,r_2)$, and $\tilde{\mathcal{A}}$ are as defined in Lemma~\ref{lem:energy_cov_UL}, and $\mathcal{L}_{\tilde{I}_2}(s)$ is defined in Lemma~\ref{lem:UL_SINR_simult}.
\end{theorem}
\begin{IEEEproof}
See Appendix~\ref{app:DL}.
\end{IEEEproof}
By comparing the above result with the uplink coverage probability of the regularly powered network given by Eq.~\ref{14}, we note that the effect of energy harvesting mainly appears in the term $\tilde{\mathcal{A}}$. For instance, if we try to exclude the energy coverage condition ($E_{\rm H}\geq \tau_3T\rho\|x_1\|^{\epsilon\alpha}$) by putting $\tau_3=0$, we will get $\tilde{C}(\tau_1)=0$, which will tend $\tilde{\mathcal{A}}$ to $\infty$. This will eventually make all the terms in Eq.~\ref{24} tend to zero except the first term which will be equivalent to Eq.~\ref{14}.



\begin{remark} \label{rem:Psuc3}
Similar to Remark~\ref{rem:Pcov2}, the value of $\tilde{\mathcal{A}}$ here represents a threshold on the distance to the second nearest BS $r_2$. In particular, as long as $r_2\leq\tilde{\mathcal{A}}$, the uplink coverage probability of the RF-powered network is exactly the same as that of the regularly powered network. This can be deduced from the first term of Eq.~\ref{24}. 
\end{remark}
Similar to Eq.~\ref{18}, the average uplink throughput $D_{\rm avg}^{\rm UL}=\tau_3R_{\rm avg}^{\rm UL}$ can be expressed as
\begin{align}
\label{25}
D_{\rm avg}^{\rm UL}= \tau_3W_U\log(1+\beta_{\rm UL})P_{\rm suc}^{\rm UL}.
\end{align}
Note that, similar to Remark~\ref{rem:Ravg}, the time-slot division parameter $\tau_1$ has an optimum value that maximizes the throughput $D_{\rm avg}^{\rm UL}$. We conclude this Section with the following remark.
\begin{remark}
By comparing the results for downlink and uplink modes (given in Theorems~\ref{thm:Pcov} and~\ref{thm:Psuc}, respectively), with those of the regularly powered network, we conclude that $\mathcal{A}=\left(\frac{2\pi\lambda_{b}}{C(\tau_1)(\alpha-2)}\right)^\frac{1}{\alpha-2}$ and $\tilde{\mathcal{A}}=\left(\frac{2\pi\lambda_{b}}{\tilde{C}(\tau_1)(\alpha-2)}\right)^\frac{1}{(\epsilon+1)\alpha-2}$ can be used as tuning parameters for the energy harvesting network. The closer we need the downlink or uplink performance to be to the regularly powered network, the larger the values of $\mathcal{A}$ and $\tilde{\mathcal{A}}$ need to be. These tuning parameters capture in their definitions the effects of all system parameters including $P_{\rm t}$, $\lambda_b$, $\tau_1$, $\tau_2$, $\tau_3$, and $\eta$.
\end{remark}
\section{Simulation Results and Discussion}\label{sec:sim}
\begin{figure}
\centering
\includegraphics[width=0.48\columnwidth]{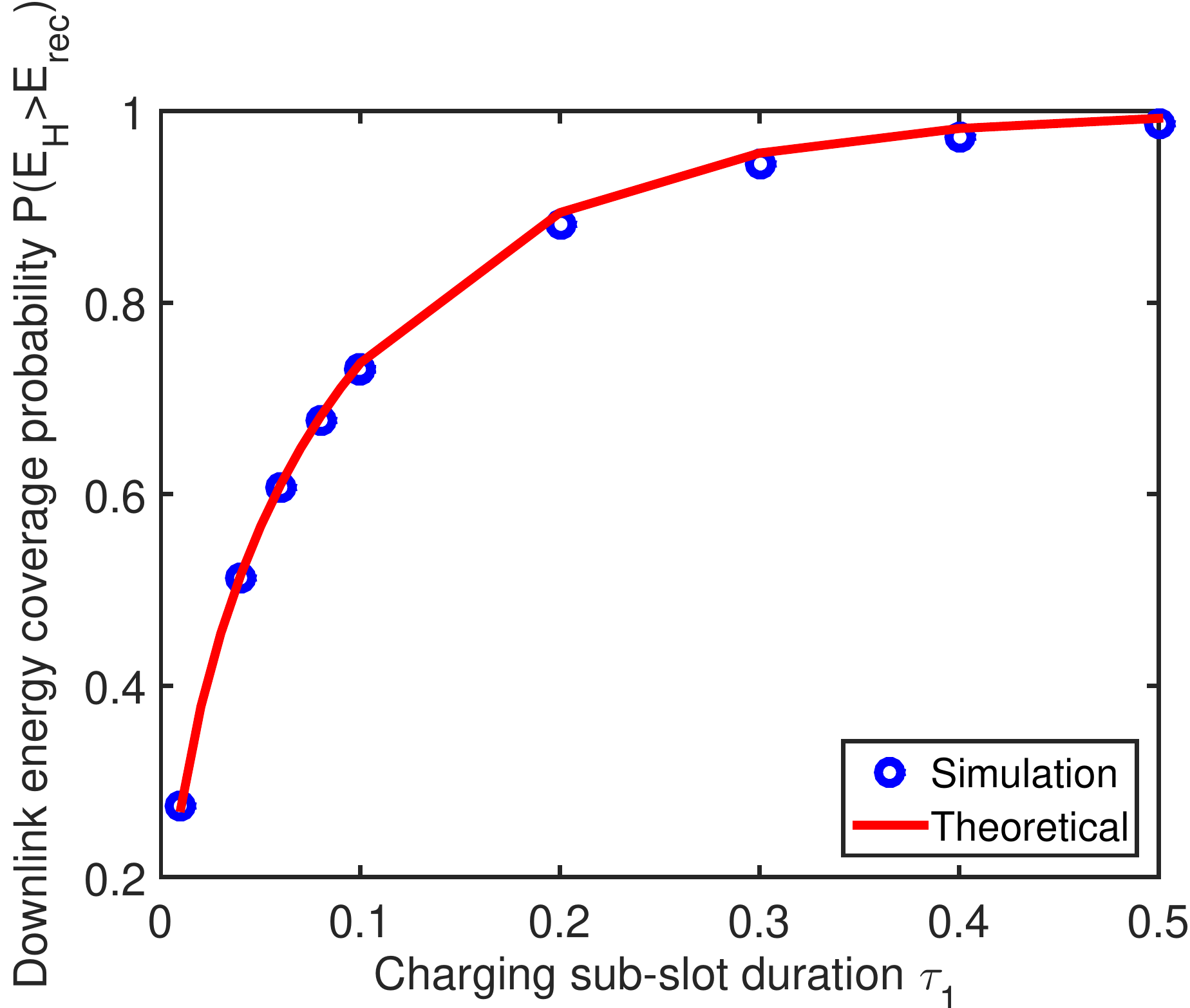}
\caption{Energy coverage probability in the downlink mode as a function of $\tau_1$.}
\label{fig:DL_energy}
\end{figure}
\begin{figure}
\centering
\includegraphics[width=0.48\columnwidth]{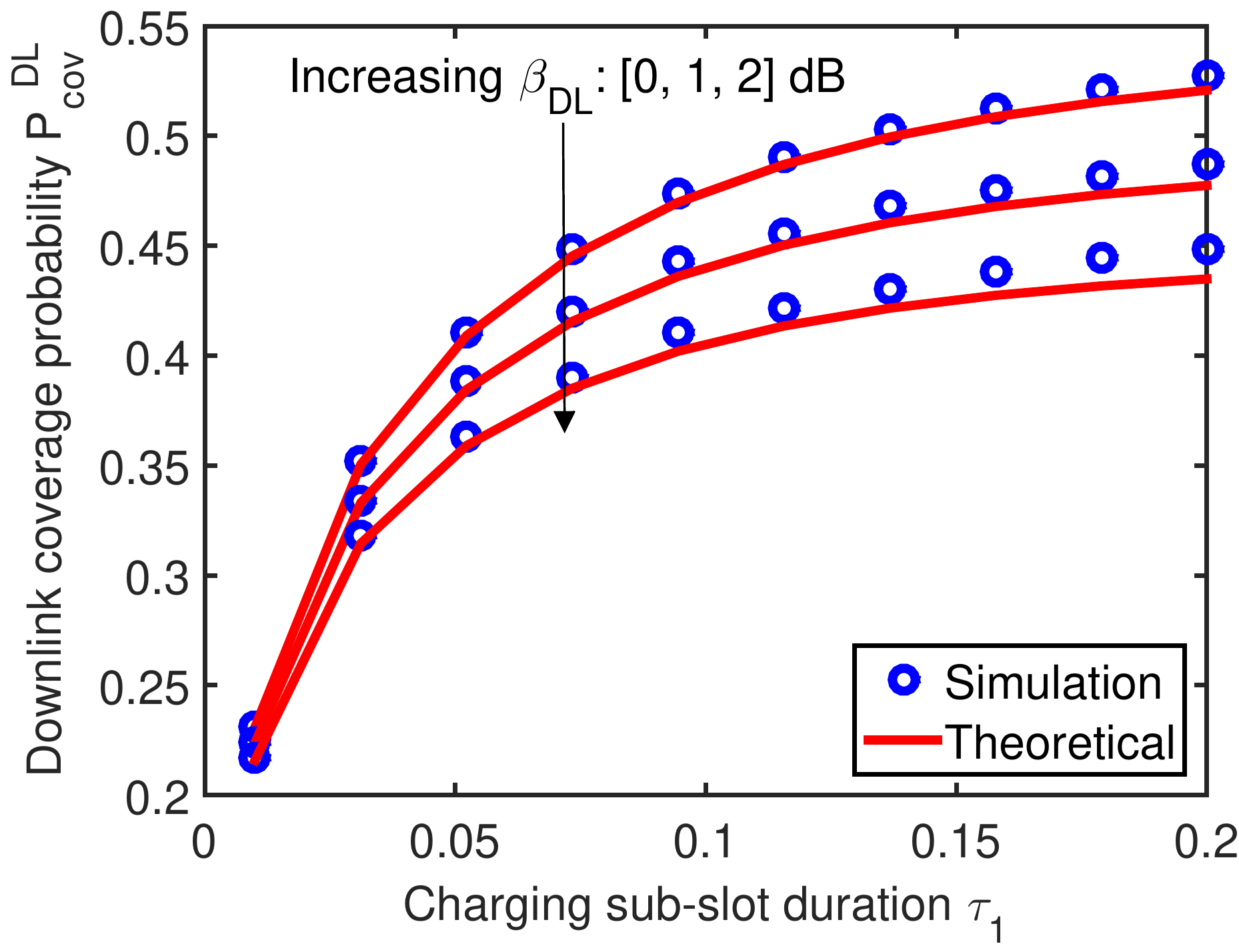}
\caption{Downlink coverage probability $P_{\rm cov}^{\rm DL}$ in the downlink mode as a function of $\tau_1$.}
\label{fig:DL_cov}
\end{figure}
\begin{figure}
\centering
\includegraphics[width=0.48\columnwidth]{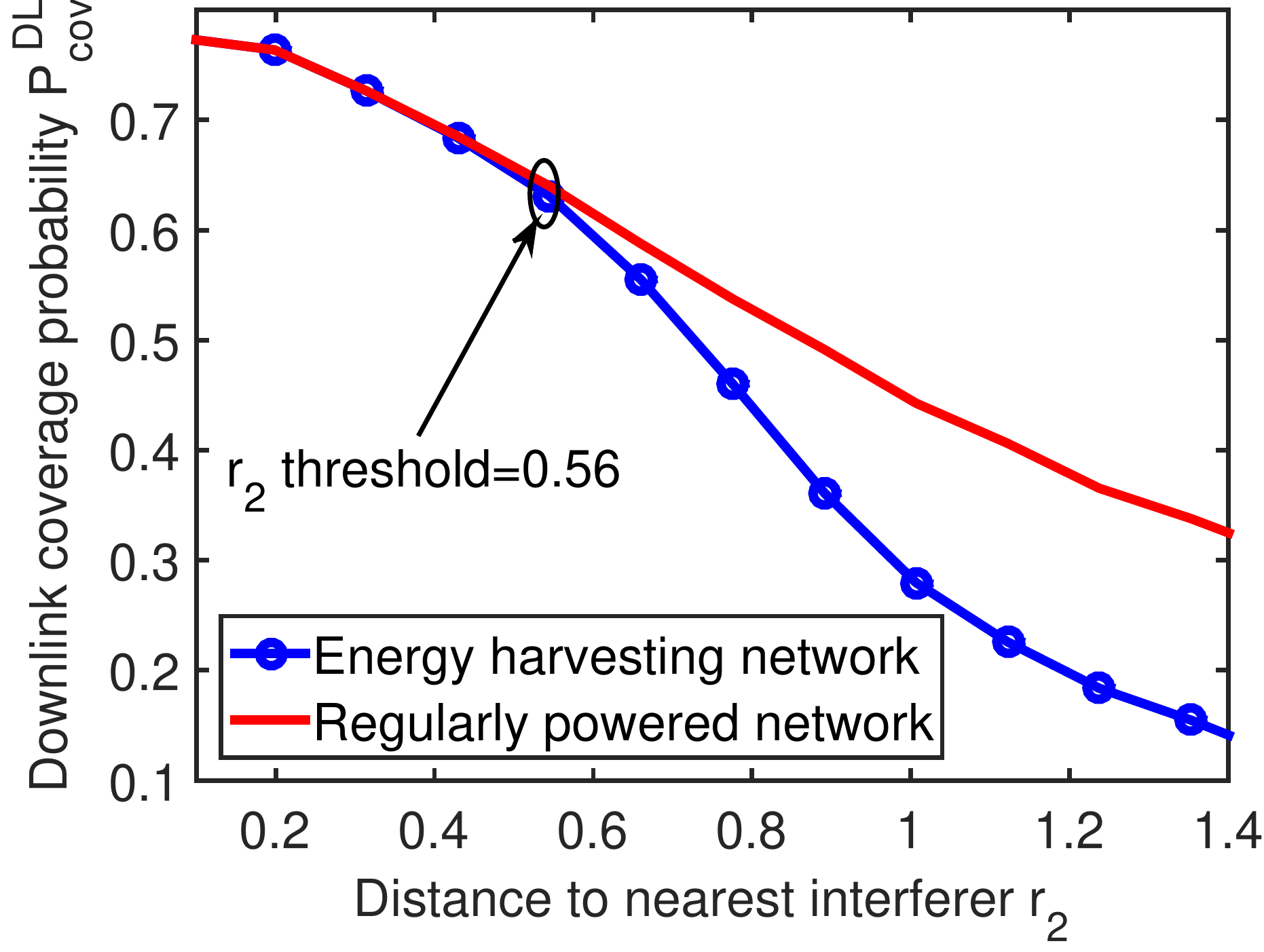}
\caption{Downlink coverage probability conditioned on the value of $r_2$.}
\label{fig:DL_threshold}
\end{figure}
\begin{figure}
\centering
\includegraphics[width=0.48\columnwidth]{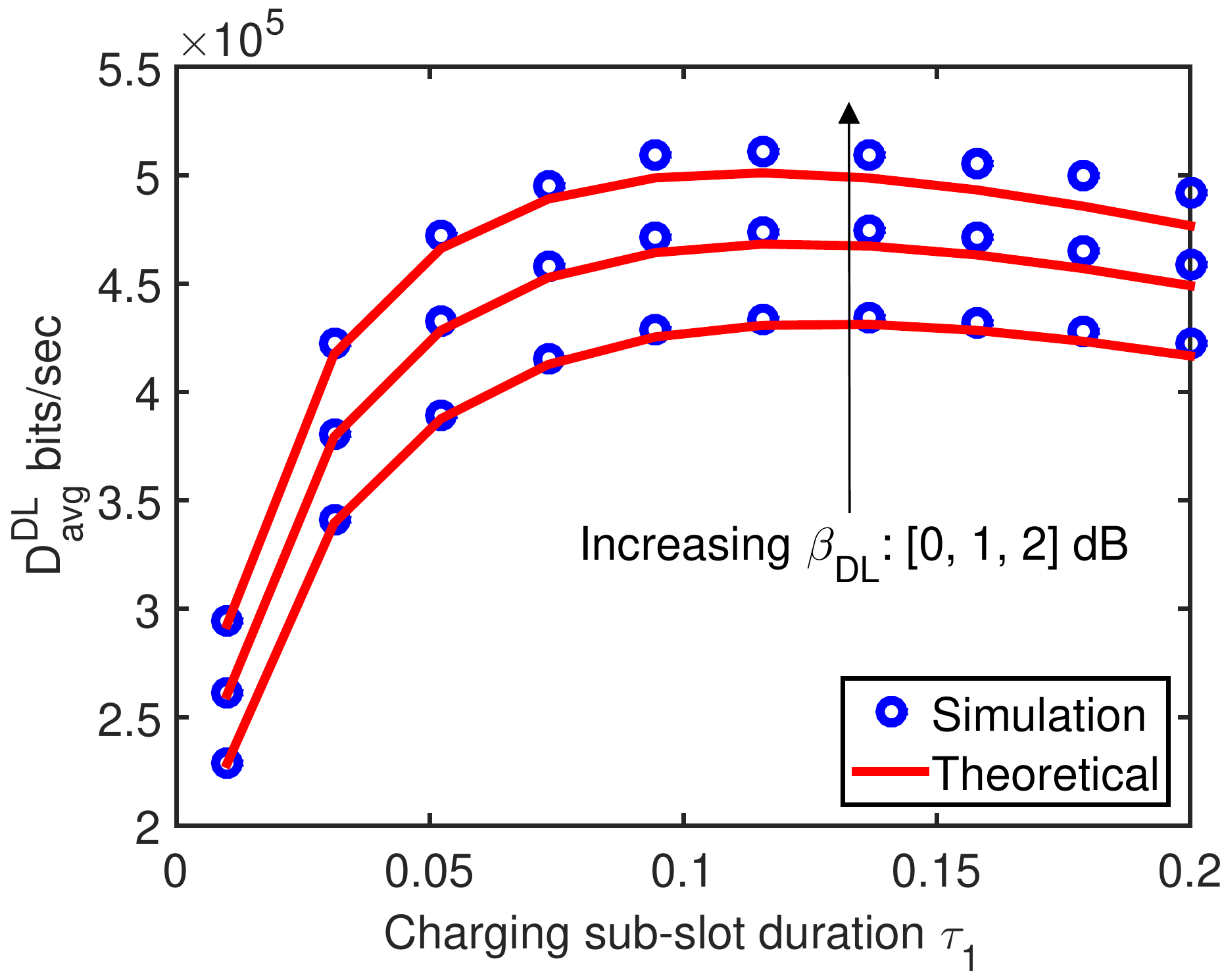}
\caption{Downlink average throughput $D_{\rm avg}^{\rm DL}$ in the downlink mode as a function of $\tau_1$.}
\label{fig:DL_Davg}
\end{figure}
Unless specified otherwise, we will consider the following values for the simulation parameters throughout this section: $E_{\rm rec}=10^{-5}$ Joules, $\lambda_{b}=1$, $\alpha=4$, $\eta=10^{-3}$, $W_D=1$ MHz, $\beta_{\rm DL}=1$ dB, $P_{\rm t}=0$ dB, $\frac{P_t}{\sigma^2_{\rm DL}}=20$ dB, $\rho=1$ dBm, $\frac{\rho}{\sigma^2_{\rm UL}}=20$ dB, $\lambda_u=30\lambda_b$, $\beta_{\rm UL}=1$ dB, $\epsilon=0.8$, and $T=10^{-2}$ sec.
\subsection{Downlink Mode}
In this subsection, we evaluate the performance of the downlink mode using the performance metrics derived in Section~\ref{sec:DL}. First, in Fig.~\ref{fig:DL_energy} we plot the energy coverage probability result derived in Lemma~\ref{lem:energy_cov}. As discussed in Remark~\ref{rem:energy_cov}, energy coverage probability is clearly an increasing function of the time division parameter $\tau_1$. The theoretical results are also shown to match perfectly with the simulation results obtained from Monte-Carlo trials, which verifies the accuracy of the dominant BS-based approach used to approximate the energy $E_{\rm H}$ in our analysis. The downlink coverage result derived in Theorem~\ref{thm:Pcov} is plotted in Fig.~\ref{fig:DL_cov}. Comparisons with simulation results again verify the accuracy of the dominant BS-based approximation. As discussed in Remark~\ref{rem:Pcov}, we notice that the coverage probability $P_{\rm cov}^{\rm DL}$ starts converging to the coverage probability of regularly powered networks, given by $\mathbb{P}({\rm SINR_{DL}}\geq\beta_{\rm DL})$, at high values of $\tau_1$. To glean sharper insights, we recall Remark~\ref{rem:Pcov2}, where we referred to $\mathcal{A}$ as a threshold on the value of distance to the second nearest BS $r_2$, below which this RF-powered IoT network has the same downlink coverage as the regularly powered network. In Fig.~\ref{fig:DL_threshold}, we verify this insight by plotting the coverage probabilities for both RF-powered and regularly powered networks conditioned on $r_2$ (for $\tau_1=0.1$). As predicted in Remark~\ref{rem:Pcov2}, the performance of both the networks is the same when $r_2$ is below the threshold value, which in this case is $r_2=\mathcal{A}=0.56$. Even thought this insight was a byproduct of dominant BS-based approximation, we notice that it is remarkably accurate. Right after the threshold value of $r_2=\mathcal{A}=0.56$, the two curves start diverging. Finally, we plot our results in Eq.~\ref{22} for the average throughput in Fig.~\ref{fig:DL_Davg}. Comparisons with the simulation results verify the accuracy of our analysis. The results also illustrate the existence of an optimum value for $\tau_1$ that maximizes the average throughput in the downlink mode, as predicted in Remark~\ref{rem:Ravg}.
\subsection{Uplink Mode}
\begin{figure}
\centering
\includegraphics[width=0.48\columnwidth]{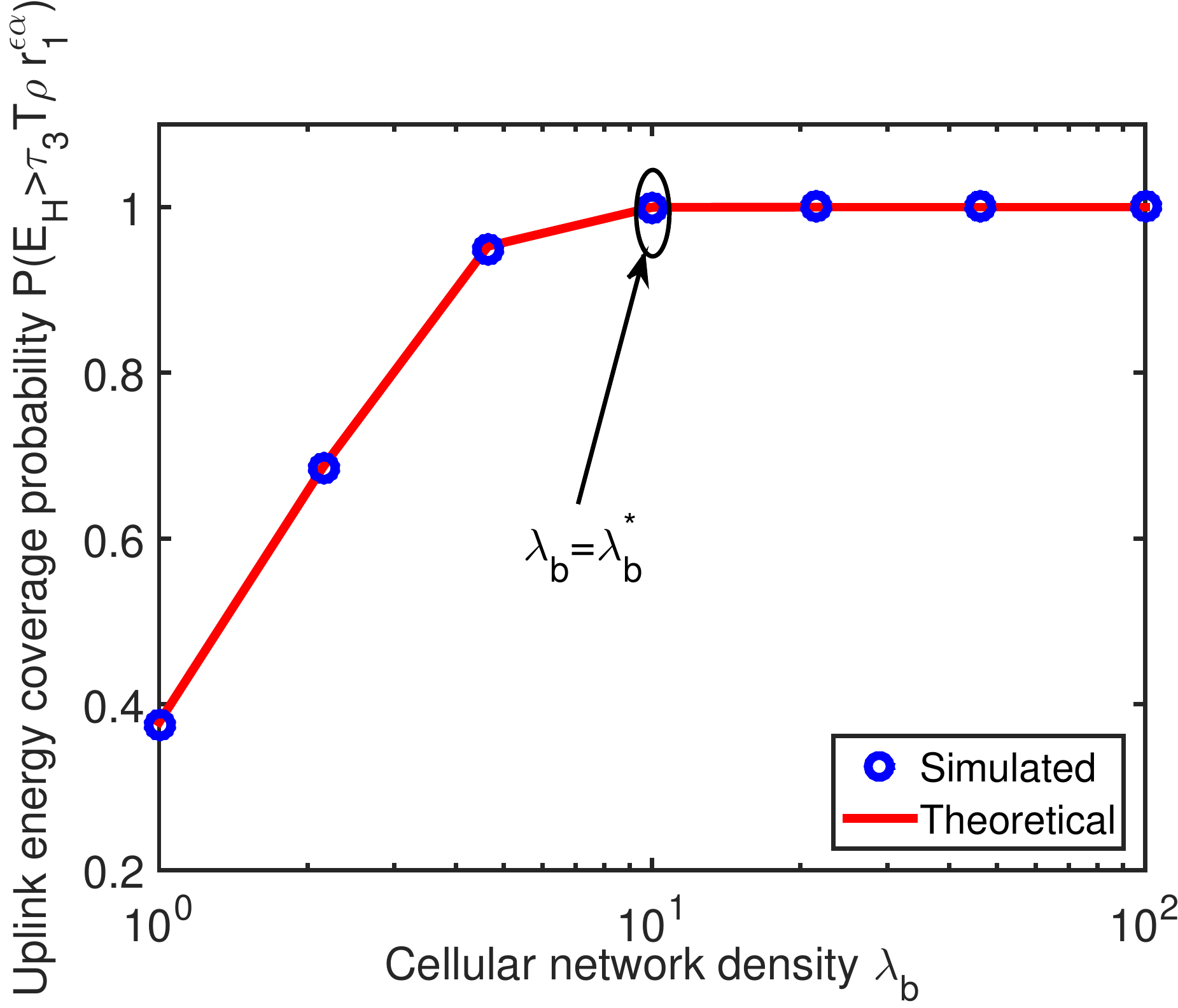}
\caption{Uplink energy coverage probability as a function of cellular network density $\lambda_{b}$.}
\label{fig:UL_energy_lambda}
\end{figure}

\begin{figure}
\centering
\includegraphics[width=0.48\columnwidth]{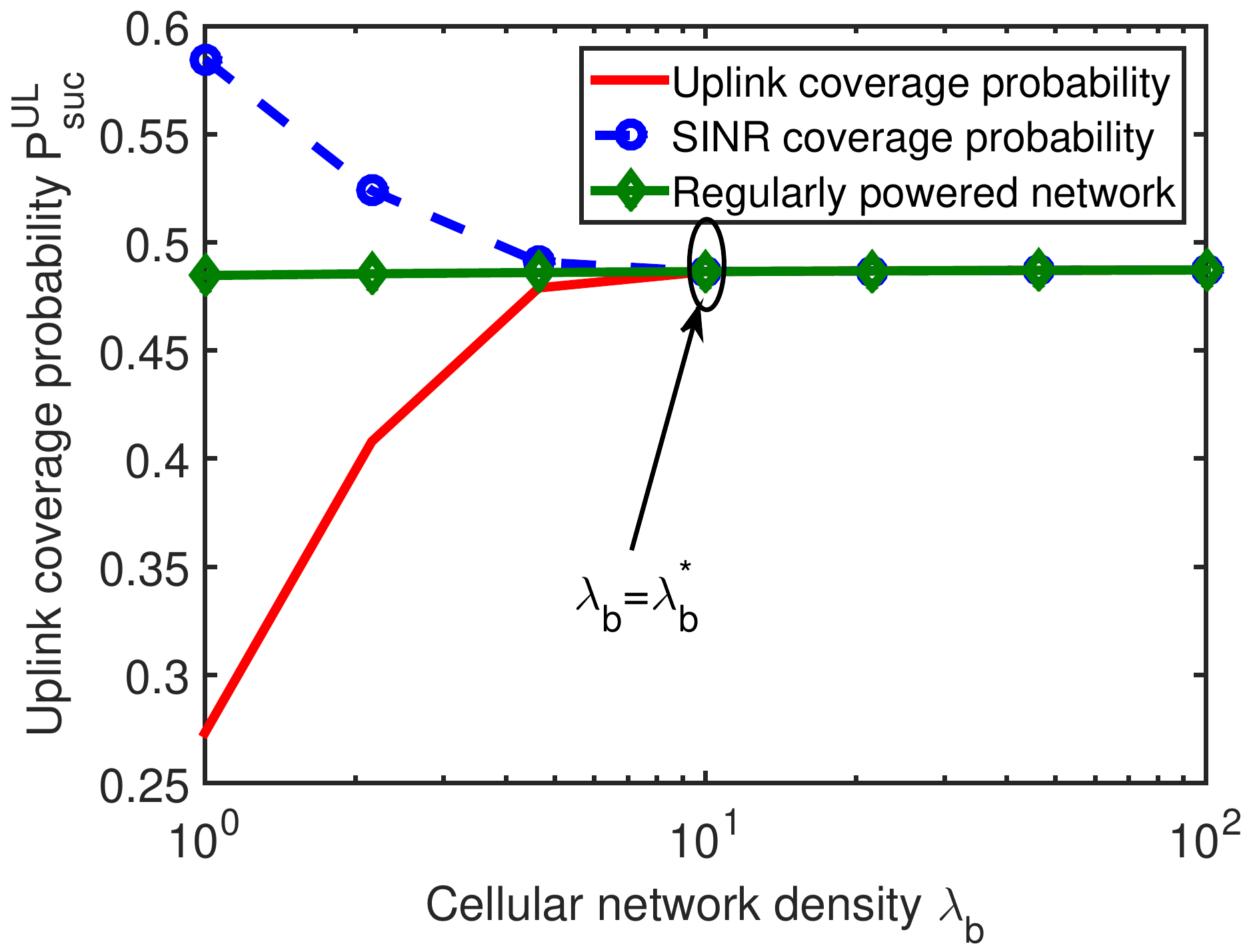}
\caption{Uplink coverage probability $P_{\rm suc}^{\rm UL}$ in the uplink mode as a function of $\lambda_{b}$.}
\label{fig:UL_Psuc_lambda}
\end{figure}

\begin{figure}
\centering
\includegraphics[width=0.48\columnwidth]{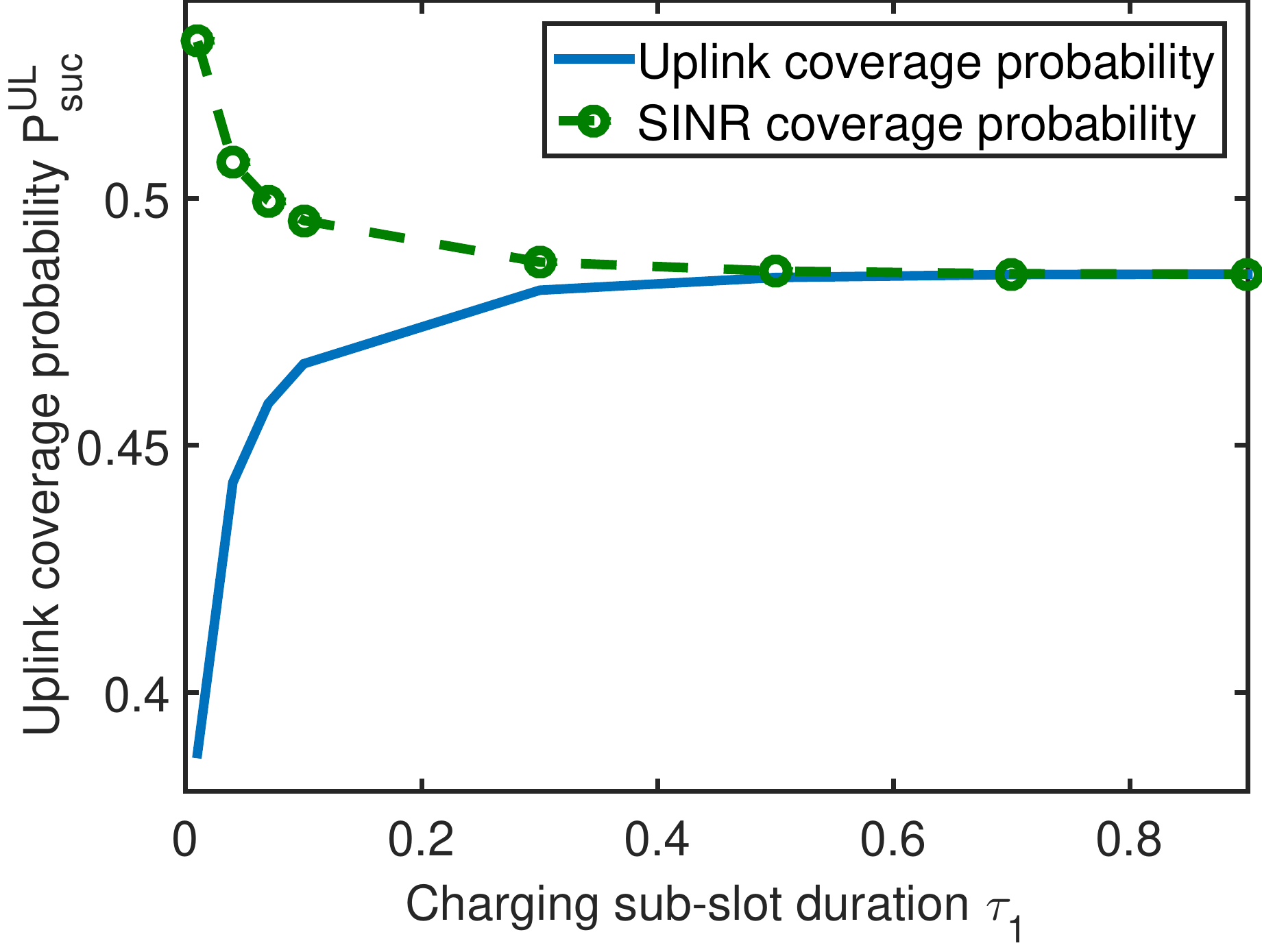}
\caption{Uplink coverage probability as a function of uplink time-slot division parameter $\tau_1$.}
\label{fig:UL_Psuc_kappa}
\end{figure}

In this section, we focus on the performance analysis of uplink mode. In particular, we will study the effect of $\tau_1$ and $\lambda_{b}$ on the performance metrics derived in Section~\ref{sec:UL}. In Fig.~\ref{fig:UL_energy_lambda}, we plot the energy coverage probability in the uplink mode as a function of $\lambda_{b}$. Consistent with Remark~\ref{rem:energy_cov_UL}, the energy coverage probability increases with $\lambda_b$ and saturates to unity when $\lambda_b$ is above a specific value, which we denote by $\lambda_b^*$. Beyond this value of density, the energy coverage condition is satisfied with high probability. Consequently, the uplink coverage probability $P_{\rm suc}^{\rm UL}$ is expected to converge to the SINR coverage probability, defined as $\mathbb{P}({\rm SINR_{UL}}\geq\beta_{\rm UL})$, at $\lambda_b^*$. This is verified in Fig.~\ref{fig:UL_Psuc_lambda}, where starting from $\lambda_{b}=\lambda_b^{*}$, the energy coverage condition is satisfied most of the time and the uplink coverage probability reduces to SINR coverage, i.e., $\mathbb{P}({\rm SINR_{UL}}\geq\beta_{\rm UL},E_{H}\geq \rho r_1^{\epsilon\alpha})\simeq \mathbb{P}({\rm SINR_{UL}}\geq\beta_{\rm UL})$. We also note that the SINR coverage probability in Fig.~\ref{fig:UL_Psuc_lambda} initially decreases with $\lambda_{b}$ until it becomes constant starting from about $\lambda_{b}=\lambda_b^*$. This is due to the increase in energy coverage probability which leads to increase in the density of active devices, hence increasing the interference value. The value to which they converge starting from $\lambda_{b}=\lambda_b^*$ is the uplink coverage probability for the case of regularly powered network ($P_{\rm suc}^{\rm UL,RP}$ in Eq.~\ref{14}). Similar trends are observed in Fig.~\ref{fig:UL_Psuc_kappa}, where we note that the uplink coverage and the SINR coverage probabilities converge at about $\tau_1=0.5$, which can be interpreted as the minimum value of $\tau_1$ at which the energy coverage condition is satisfied with a high probability. Also, similar to our discussion above on the effect of $\lambda_{b}$, the SINR coverage probability in Fig.~\ref{fig:UL_Psuc_kappa} initially decreases due to the increase in the energy coverage probability which increases the density of active devices and, consequently, the interference.


\subsection{Joint Uplink and Downlink Mode}
\begin{figure}
\centering
\includegraphics[width=0.48\columnwidth]{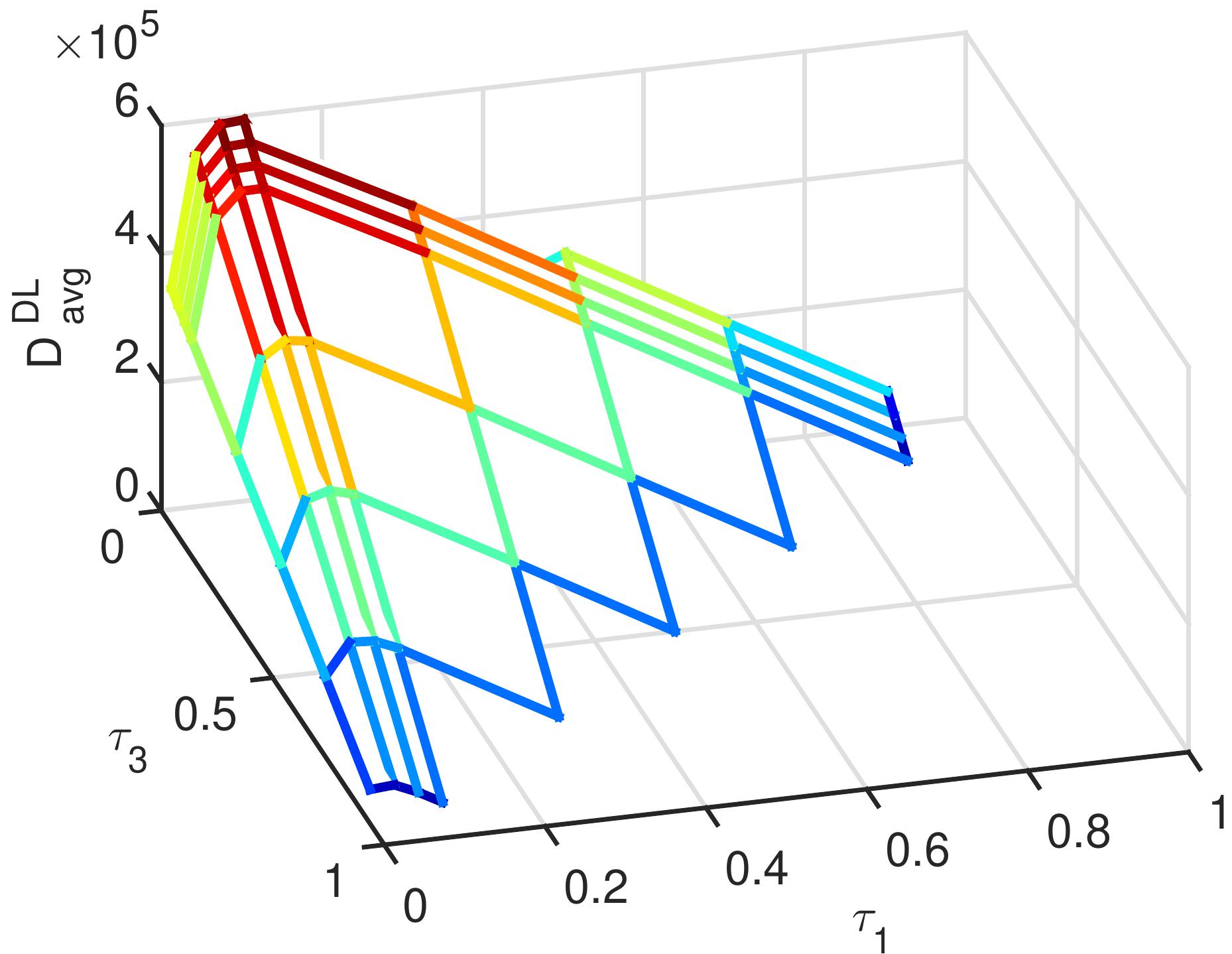}
\caption{Downlink average throughput during joint uplink and downlink mode as a function of $\tau_1$ and $\tau_3$.}
\label{fig:simult_DL}
\end{figure}

In Fig.~\ref{fig:simult_DL} we provide a 3D plot for $D_{\rm avg}^{\rm DL}$ as a function of $\tau_1$ and $\tau_3$. Recall that $\tau_2 = 1 - \tau_1 - \tau_3$. We note that for any given value of $\tau_1$, the value of $D_{\rm avg}^{\rm DL}$ decreases as $\tau_3$ increases (equivalently $\tau_2$ decreases). As discussed in Remark~\ref{rem:Ravg_simult}, for any given value of $\tau_3$, there exists optimal $\tau_1$ (and hence optimal $\tau_2$) that maximizes $D_{\rm avg}^{\rm DL}$. A similar behavior has already been seen in Fig.~\ref{fig:DL_Davg}. Similar observations can be made about the behavior of $D_{\rm avg}^{\rm UL}$.
\section{Conclusions}
In this paper, we developed an analytical framework to study joint uplink and downlink coverage performance of a cellular-based ambient RF energy harvesting network in which IoT devices are solely powered by the downlink cellular transmissions. Each time-slot is assumed to be partitioned into charging, downlink, and uplink sub-slots. Within each time-slot, the IoT devices (assumed batteryless) first harvest RF energy from cellular transmissions and then use this energy to perform downlink and uplink communication in the subsequent sub-slots. For this setup, we derived the joint probability that the typical device harvests sufficient energy in the charging sub-slot and achieves sufficiently high downlink and uplink SINRs in the following two sub-slots. The main technical contribution is in handling the correlation between these energy and SINR coverage events. Using this result, we also studied system throughput as a function of the time-slot division parameters. Optimal slot partitioning that maximizes this throughput is also discussed. Using these results, we also compared the performance of this RF-powered IoT network with a regularly powered network in which the IoT devices have uninterrupted access to reliable power source, such as a battery. We derived thresholds on several system parameters beyond which the performance of this RF-powered IoT network converges to that of the regularly powered network. 

Finally, we defined a {\em tuning parameter}, which incorporates the effect of all system parameters, and needs to be sufficiently high for the coverage performance of this RF-powered network to converge to that of the regularly powered network.

This work can be extended in multiple directions. From the energy harvesting perspective, the system model can be extended to include rechargeable batteries (with finite capacities) at the devices. This will require explicit consideration of the temporal dimension, as done in \cite{DhiLiJ2014}, where the BSs were assumed to be self-powered with access to batteries with finite capacities. From the modeling perspective, it is important to consider other BS-device configurations, such as the ones in which devices are clustered around the BSs~\cite{SahAfsJ2017}. 

\appendices
\section{}\label{app:energy_cov_simult}
The value of $\Psi(r_2)$ can be derived as follows:
\begin{align}
\label{26}
\Psi(r_2) = \mathbb{E}\left[\sum_{x\in \Phi_{b}\backslash{x_1,x_2}}g_x\|x\|^{-\alpha}\Big|x_1,x_2\right] &\stackrel{(a)}{=} \mathbb{E}\left[\displaystyle\sum_{x_i\in\Phi_{b}\backslash x_1,x_2}\|x_i\|^{-\alpha}\right] \stackrel{(b)}{=} 2\pi\lambda_{b}\displaystyle\int_{r_2}^{\infty}\frac{1}{r^\alpha}r{\rm d}r=\frac{2\pi\lambda_{b}}{\alpha-2}({r_2}^{2-\alpha}),
\end{align}
where (a) follows from the assumption that all $\{g_{x}\}$ are independent and exponentially distributed random variables with mean one, and (b) follows from Campbell's theorem~\cite{haenggi2012stochastic} with conversion from Cartesian to polar coordinates and using $r_2=\|x_2\|$. Using the approximation introduced in Eq.~\ref{9}, the conditional energy coverage probability can be expressed as:
\begin{align}
\label{27}
\mathbb{P}&\left(E_{\rm H}\geq E_{\rm min}\Big|\Phi_{b}\right)=\mathbb{P}\left(\tau_1 T\eta P_{\rm t}\left(g_{x_1}r_1^{-\alpha}+g_{x_2}r_2^{-\alpha}+\frac{2\pi\lambda_{b}}{\alpha-2}r_2^{2-\alpha}\right)\geq E_{\rm rec}+\tau_3 T\rho r_1^{\epsilon\alpha}\right)\nonumber\\
&=\mathbb{P}\left(g_{x_1} r_1^{-\alpha}+g_{x_2} r_2^{-\alpha}\geq C(\tau_1)+\frac{\tau_3 \rho r_1^{\epsilon\alpha}}{\tau_1 \eta P_{\rm t}}-\frac{2\pi\lambda_{b}}{\alpha-2}r_1^{2-\alpha}\right)=\mathbb{P}\left(g_{x_1} r_1^{-\alpha}+g_{x_2} r_2^{-\alpha}\geq \mathcal{F}(r_1,r_2)\right)\nonumber\\
&\stackrel{(c)}{=}\frac{r_2^{\alpha}\exp(-r_1^{\alpha}[\mathcal{F}(r_1,r_2)]^{+})-r_1^{\alpha}\exp(-r_2^{\alpha}[\mathcal{F}(r_1,r_2)]^{+})}{r_2^{\alpha}-r_1^{\alpha}},
\end{align}
where step (c) is due to hypo-exponential distribution of $g_{x_1} r_1^{-\alpha}+g_{x_2} r_2^{-\alpha}$ (sum of two exponential random variables with rates $r_1^{\alpha}$ and $r_2^{\alpha}$), $C(\tau_1)=\frac{E_{\rm rec}}{\tau_1 T\eta P_{\rm t}}$, and $[x]^{+}=\max\{0,x\}$. This concludes the proof of Eq.~\ref{10}. Noting that $\mathbb{P}\left(E_{\rm H}\geq E_{\rm min}\Big|\Phi_{b}\right)=1$ when $\mathcal{F}(r_1,r_2)\leq 0$ and integrating over $r_1$ and $r_2$ with $f_{R_1,R_2}(r_1,r_2)=(2\pi\lambda_{b})^{2} r_1r_2e^{-\lambda_{b}\pi {r_2}^2}$~\cite{moltchanov2012distance}, the result in Eq.~\ref{11} follows.
\section{}\label{app:SINR_simult}
Using the definition of ${\rm SINR_{DL}}$ in Eq.~\ref{1} and approximating the interference $I_1$ by the sum of interference from the nearest interferer and the expectation of the interference from the rest of the interference field, we get
\begin{align}
\label{28}
\mathbb{P}&({\rm SINR_{DL}}\geq {\beta_{\rm DL}}|r_1,r_2) = \mathbb{P}\left(\frac{P_{\rm t}h_{x_1}\|x_1\|^{-\alpha}}{I_1+\sigma_{\rm DL}^2}\geq {\beta_{\rm DL}}\Big|r_1,r_2\right)
\nonumber\\&=\mathbb{P}\left(\frac{P_{\rm t}h_{x_1}r_1^{-\alpha}}{P_{\rm t}h_{x_2}r_2^{-\alpha}+P_{\rm t}\Psi(r_2)+\sigma_{\rm DL}^2}\geq {\beta_{\rm DL}}\Big|r_1,r_2\right)\stackrel{(d)}{=} \mathbb{P}\left(\frac{P_{\rm t}h_{x_1}r_1^{-\alpha}}{P_{\rm t}h_{x_2}r_2^{-\alpha}+P_{\rm t}\frac{2\pi\lambda_br_2^{2-\alpha}}{\alpha-2}+\sigma_{\rm DL}^2}\geq {\beta_{\rm DL}}\Big|r_1,r_2\right)\nonumber\\
&= \mathbb{P}\left(h_{x_1}r_1^{-\alpha}\geq\frac{\beta_{\rm DL}\sigma_{\rm DL}^2}{P_{\rm t}}+\frac{2\pi\lambda_b\beta_{\rm DL}r_2^{2-\alpha}}{\alpha-2}+\beta_{\rm DL}h_{x_2}r_2^{-\alpha}\right)\\
&\stackrel{(e)}{=}\mathbb{E}_{h_{x_2}}\left[\exp\left(-r_1^{\alpha}\left(\frac{\beta_{\rm DL}\sigma_{\rm DL}^2}{P_{\rm t}}+\frac{2\pi\lambda_b\beta_{\rm DL}r_2^{2-\alpha}}{\alpha-2}+\beta_{\rm DL}h_{x_2}r_2^{-\alpha}\right)\right)\right]\stackrel{(f)}{=}\exp(-\mathcal{G}(r_1,r_2))\frac{1}{1+\beta_{\rm DL}\frac{r_1^{\alpha}}{r_2^{\alpha}}},\nonumber
\end{align}
where (d) follows from substituting for $\Psi(r_2)$ as derived in Eq.~\ref{26}, and steps (e) and (f) follow from the assumption that $h_{x}\sim\exp(1)$, and defining $\mathcal{G}(r_1,r_2)=\frac{\beta_{\rm DL}\sigma_{\rm DL}^2r_1^{\alpha}}{P_{\rm t}}+\frac{2\pi\lambda_b\beta_{\rm DL}r_2^{2-\alpha}r_1^{\alpha}}{\alpha-2}$.

In the uplink sub-slot, the locations of {\em active} IoT devices (IoT devices in energy coverage) in a given time-frequency resource can be approximately modeled by the PPP $\tilde{\Phi}_u$ with density $\tilde{\lambda}_u=P_h\times\lambda_b$ where $P_h=\mathbb{P}(E_{\rm H}\geq E_{\rm min})$. This will lead to the following expression for ${\rm SINR_{UL}}$:
\begin{align}
\label{29}
{\rm SINR_{UL}}=\frac{w_o\|x_1\|^{(\epsilon-1)\alpha}}{\sum\limits_{u_i\in\tilde{\Phi}_u\backslash u_o} w_i \left(R_1^{(i)}\right)^{\epsilon\alpha}D_i^{-\alpha}+\frac{\sigma_{\rm UL}^2}{\rho}}.
\end{align}
Defining $\tilde{I}_2=\sum\limits_{u_i\in\tilde{\Phi}_u\backslash u_o} w_i \left(R_1^{(i)}\right)^{\epsilon\alpha}D_i^{-\alpha}$, we have:
\begin{align}
\label{30}
\mathbb{P}&\left({\rm SINR_{UL}}\geq \beta_{\rm UL}\Big|r_1\right) = \mathbb{P}\left(\frac{w_0{r_1}^{(\epsilon-1)\alpha}}{\tilde{I}_2+\frac{\sigma_{\rm UL}^2}{\rho}}\geq \beta_{\rm UL}\Big|r_1\right)=\mathbb{E}_{\tilde{I}_2}\left[\mathbb{P}\left(w_0\geq \frac{{(\tilde{I}_2+\frac{\sigma_{\rm UL}^2}{\rho})}\beta_{\rm UL}}{{r_1}^{(\epsilon-1)\alpha}}\Big|r_1,\tilde{I}_2\right)\right]\nonumber\\
&\stackrel{(g)}{=}\mathbb{E}_{\tilde{I}_2}\left[\exp\left(-\frac{{(\tilde{I}_2+\frac{\sigma_{\rm UL}^2}{\rho})}\beta_{\rm UL}}{{r_1}^{(\epsilon-1)\alpha}}\right)\right]\stackrel{(h)}{=} e^{\left(-\frac{\beta_{\rm UL}\sigma_{\rm UL}^2}{\rho {r_1}^{(\epsilon-1)\alpha}}\right)}\mathcal{L}_{\tilde{I}_2}\left(\frac{\beta_{\rm UL}}{{r_1}^{(\epsilon-1)\alpha}}\right),
\end{align}
where step (g) is due to the assumption that $w_0$ is exponentially distributed with mean one, and step (h) results from using the Laplace transform of $\tilde{I}_2$, which can be found by replacing $\lambda_b$ with $\tilde{\lambda}_b=P_h \lambda_b$ in Eq.~\ref{13}, where $P_h=\mathbb{P}(E_{\rm H}\geq E_{\rm min})$.
\section{}\label{app:DL}
We apply the substitutions in Remark~\ref{rem:simult} for the downlink case to both Lemma~\ref{lem:energy_cov_simult} and Theorem~\ref{thm:simult} to get both energy coverage probability and $P_{\rm cov}^{\rm DL}$. Applying these substitutions reduces the value of $\mathcal{F}(r_1,r_2)$ to $\mathcal{F}_{\rm DL}(r_1,r_2)=C(\tau_1)-\frac{2\pi\lambda_br_2^{2-\alpha}}{\alpha-2}$, where $C(\tau_1)$ is as defined in Lemma~\ref{lem:energy_cov_simult}. Letting $\mathcal{A}=\left(\frac{2\pi\lambda_b}{(\alpha-2)C(\tau_1)}\right)^{\frac{1}{\alpha-2}}$, we note that the set $\mathcal{N}_{r_2}$ will be empty set for $r_2\geq\mathcal{A}$ while for $r_2\leq\mathcal{A}$ the set will be simply $\mathcal{N}_{r_2}=\{r_1:r_1\leq r_2\}$. Similarly, the set $\mathcal{P}_{r_2}$ will be empty set for $r_2\leq\mathcal{A}$ while for $r_2\geq\mathcal{A}$ the set will reduce to $\mathcal{P}_{r_2}=\{r_1:r_1\leq r_2\}$. Applying these integration limits on our result in Lemma~\ref{lem:energy_cov_simult} leads to the final result in Lemma~\ref{lem:energy_cov}. Similarly, applying these new integration limits to the result in Theorem~\ref{thm:simult} and noting that the substitutions explained in Remark~\ref{rem:simult} include $\beta_{\rm UL}=0$ (which leads to $\mathcal{L}_{\tilde{I}_2}(0)=1$ in Eq.~{16}), the final result in Theorem~\ref{thm:Pcov} follows.
\section{}\label{app:UL}
Similar to the approach in the downlink case, we apply the substitutions in Remark~\ref{rem:simult} for the uplink case to both Lemma~\ref{lem:energy_cov_simult} and Theorem~\ref{thm:simult}. Applying these substitutions reduces the value of $\mathcal{F}(r_1,r_2)$ to $\mathcal{F}_{\rm UL}(r_1,r_2)=\tilde{C}(\tau_1)r_1^{\epsilon\alpha}-\frac{2\pi\lambda_br_2^{2-\alpha}}{\alpha-2}$, where $\tilde{C}(\tau_1)=\frac{\tau_3\rho }{\tau_1\eta P_{\rm t}}$. Letting $\tilde{\mathcal{A}}=\left(\frac{2\pi\lambda_b}{\tilde{C}(\tau_1)(\alpha-2)}\right)^{\frac{1}{(\epsilon+1)\alpha-2}}$, we note that the set $\mathcal{N}_{r_2}=\{r_1:r_1<\left(\frac{2\pi\lambda_b}{\tilde{C}(\tau_1)(\alpha-2)}\right)^{\frac{1}{\epsilon\alpha}}r_2^{\frac{2-\alpha}{\epsilon\alpha}}\}$ for $r_2\geq\tilde{\mathcal{A}}$ while for $r_2\leq\tilde{\mathcal{A}}$ the set will be simply $\mathcal{N}_{r_2}=\{r_1:r_1\leq r_2\}$. Similarly, the set $\mathcal{P}_{r_2}$ will be empty set for $r_2\leq\tilde{\mathcal{A}}$ while for $r_2\geq\tilde{\mathcal{A}}$ the set will reduce to $\mathcal{P}_{r_2}=\{r_1:\left(\frac{2\pi\lambda_b}{\tilde{C}(\tau_1)(\alpha-2)}\right)^{\frac{1}{\epsilon\alpha}}r_2^{\frac{2-\alpha}{\epsilon\alpha}}\leq r_1\leq r_2\}$. Applying these integration limits on our result in Lemma~\ref{lem:energy_cov_simult} leads to the final result in Lemma~\ref{lem:energy_cov_UL}. Similarly, applying these new integration limits to the result in Theorem~\ref{thm:simult} and noting that the substitutions explained in Remark~\ref{rem:simult} include $\beta_{\rm DL}=0$ (which makes $\mathcal{G}(r_1,r_2)=0$ in Eq.~{16}), the final result in Theorem~\ref{thm:Psuc} follows.
\ifCLASSOPTIONcaptionsoff
  \newpage
\fi

{ \setstretch{1.3}
\bibliographystyle{IEEEtran}
\bibliography{Draft_v0.12.bbl}
}

\end{document}